\documentclass[twocolumn, numberedappendix]{aastex62}

\usepackage{amstext}
\usepackage{amsmath}
\usepackage{amssymb}
\usepackage{graphicx}
\usepackage{upgreek}
\newcommand{\beq}{\begin{equation}}
\newcommand{\eeq}{\end{equation}}
\newcommand{\beqar}{\begin{align}}
\newcommand{\eeqar}{\end{align}}

\shorttitle{}

\shortauthors{}

\begin{document}

\title{Polygram Stars: Resonant Tidal Excitation of Fundamental Oscillation Modes in Asynchronous Stellar Coalescence}

\author[0000-0002-1417-8024]{Morgan MacLeod}
\altaffiliation{NASA Einstein Fellow}
\affiliation{Harvard-Smithsonian Center for Astrophysics, 60 Garden Street, Cambridge, MA, 02138, USA}
\email{morgan.macleod@cfa.harvard.edu}

\author[0000-0002-3752-3038]{Michelle Vick}
\affiliation{Cornell Center for Astrophysics and Planetary Science, Department of Astronomy, Cornell University, Ithaca, NY 14853, USA}

\author[0000-0002-1934-6250]{Dong Lai}
\affiliation{Cornell Center for Astrophysics and Planetary Science, Department of Astronomy, Cornell University, Ithaca, NY 14853, USA}

\author[0000-0001-5603-1832]{James M. Stone}
\affiliation{Department of Astrophysical Sciences, Princeton University, Princeton, NJ 08544, USA}

\begin{abstract}
The prevalence of binary stars at close separations implies that many of these systems will interact or merge during the binary's lifetime. This paper presents hydrodynamic simulations of the scenario of binary coalescence through unstable mass transfer, which drives the pair to closer separations. When the donor star does not rotate synchronously with respect to the orbit, dynamical tidal waves are excited in its envelope. We show that resonance crossings with high azimuthal-order $(m\sim3$ to $6$) fundamental modes induce a visible ``polygram" distortion to the star. As the binary orbit tightens, the system sweeps through resonance with modes of decreasing azimuthal order, which are selectively excited. We compare our hydrodynamic simulations to predictions from linear theory of resonant mode excitation. The linear theory provides an estimate of mode amplitudes to within a factor of  two, even as the oscillations become quite non-linear as the stars coalesce. 
We estimate that a wave with 10\% radial amplitude generates approximately 1\% photometric variability; this may be detectible if such a binary coalescence is caught in action by future photometric all-sky surveys. 
\end{abstract}


\section{Introduction} 

Approximately half of all stars reside in binary or multiple systems \citep[e.g.][]{2013ARA&A..51..269D,2017ApJS..230...15M}. Many of these systems have sufficiently small separations that stellar evolution or dynamical interactions within a multiple-star system can drive them into direct interaction \citep{2017PASA...34....1D}. These interactions can lead to mass exchange, orbital transformation, or even the complete merger of the binary pair. 

In this paper, we study a particular scenario of such hydrodynamic binary interaction: unstable mass transfer from a Roche-lobe filling donor star onto a more compact, but less massive, accretor star. This mass transfer is unstable in that the exchange of orbital angular momentum via mass removal from the donor star drives the binary system to increasing degrees of Roche lobe overflow and escalating mass transfer rate. This scenario inevitably leads to a common envelope phase in which the donor engulfs the accretor within its extended envelope \citep{1976IAUS...73...75P}. From this point, the binary may go on to merge into a single object, or to eject the donor's envelope, leaving behind a transformed binary system \citep{1993PASP..105.1373I,2010NewAR..54...65T,2013A&ARv..21...59I}. 

Many systems are thought to have evolutionary histories such that when they reach the point of Roche lobe overflow, the donor star's envelope is likely to be in synchronous rotation with the orbital motion \citep{2014ARA&A..52..171O}. In this case, the gravity of the accretor star raises a static tidal distortion of the donor star's envelope in the corotating frame of the binary. The dynamics of these synchronous scenarios have been analyzed recently by \citet{2018ApJ...863....5M,2018arXiv180805950M}. Deviation from synchronized rotation can lead to work done as the binary potential raises dynamical waves in the gaseous envelope. Wave motions, in turn, lead to dissipation, which tends to bring the system back toward the quasi-equilibrium configuration.  

However, there are several cases in which we might imagine that the synchronous configuration is never achieved. One such case, known as the Darwin tidal instability \citep{1879RSPS...29..168D}, is when there is insufficient angular momentum in the binary orbit to pull the donor's envelope into synchronous rotation \citep[e.g.][]{1973ApJ...180..307C,1980A&A....92..167H,1981A&A....99..126H}. Tidal dissipation then leads to orbital decay toward the onset of mass exchange without ever reaching a synchronous state \citep[e.g.][]{1994ApJ...423..344L,2001ApJ...562.1012E, 2017ApJ...849L..11W,2018MNRAS.481.4077S}. This scenario tends to arise when the binary mass ratio is far from unity, or when the moment of inertia of the donor's envelope is relatively large. Another situation in which synchronous rotation might not be achieved is when the mechanism that assembles the interacting binary allows insufficient time for synchronization to act. One example is in massive-star systems, where the star's evolutionary timescale can be rapid compared to the tidal dissipation timescale \citep{1983MNRAS.203..581S}.

We study the excitation of dynamical tides following the assembly of an asynchronous binary at the Roche limit, in which the donor star is beginning to lose mass. Our simulated system contains an isentropic giant-star donor interacting with a companion (``accretor") of one-tenth the donor's mass. As the orbit shrinks, the tidal forcing frequency, set by the difference between the orbital frequency and envelope rotation of the donor, passes through integer resonances with the frequencies of fundamental, acoustic modes of the star. Waves of successive azimuthal order (first high $m$ then lower $m$) are preferentially excited as the system passes through these resonances. This finding is in contrast to binaries at wider separations, which predominantly exhibit quadrupolar mode excitations by the tidal potential \cite[e.g.][]{1941MNRAS.101..367C,1994MNRAS.270..611L}. 

In Section \ref{sec:method}, we describe our simulated model and numerical approach. In Section \ref{sec:resonances}, we discuss the criteria for resonance between tidal forcing and oscillatory modes of the star. In Section \ref{sec:modes} we analyze the amplitudes resonantly-excited tidal waves that arise, and compare these results to predictions from linear theory. In Section \ref{sec:discussion}, we discuss the implications and potential observability of these oscillations, and  we conclude in Section \ref{sec:conclusion}.

\section{Numerical Model and Methods}\label{sec:method}

We report on the results of gas dynamics simulations performed using the {\tt Athena++} hydrodynamics code\footnote{Stone, J. M. et al (in preparation), online at: https://princetonuniversity.github.io/athena/}  and based on the approach developed by \citep{2018ApJ...863....5M,2018arXiv180805950M}.  Our simulated system involves a donor star of mass $M_1=1$ and radius $R_1=1$. Together with a choice of gravitational constant $G=1$, these choices define our system of units. The donor star interacts with a lower-mass accretor of mass $M_2=0.1$. Both the core of $M_1$ and the entirety of $M_2$ are treated as softened point masses. The treatment of gas self-gravity is approximate in that we apply a static potential based on the original stellar profile. We perform our calculation in the frame of the orbiting donor star, in spherical polar coordinates. We adopt an adiabatic equation of state in our models below with $\gamma=5/3$. 

We initialize the system in a circular orbit at a separation similar to the analytic Roche limit separation, at which mass exchange from the donor to the accretor is expected to commence \citep{1983ApJ...268..368E}. The donor star is initially non-spinning, and it is therefore asynchronous relative to the orbital frequency.  Mass exchange leads to accelerating orbital decay, and the system trends toward coalescence \citep{2018ApJ...863....5M}. We halt the calculation when the accretor has entered the donor's envelope,  when the binary separation is less than $R_1$. 

The donor star is initialized as a polytrope with core excised, at the origin of the spherical-polar computational mesh. The model presented here has a reflective inner boundary at $0.3R_1$, and a core mass of $0.41M_1$ inside this radius. The profile of the envelope has a structural polytropic index of $\Gamma_{\rm s}=5/3$, representative of the convective (and therefore isentropic) envelope of a red-giant star.  For a much more detailed description of the hydrodynamic approach and associated tests, we direct the reader to Section 3 and appendix A of \citep{2018ApJ...863....5M} and Section 2 of \citep{2018arXiv180805950M}. 

To examine tidal excitation, We use the package {\tt shtools} \citep{shtools}\footnote{online at: https://shtools.oca.eu/shtools/} to compute the real spherical harmonic coefficients of the  radial velocity field at the original donor-star radius according to,
\beq
v_{lm}=\frac{1}{4\pi}\int_{\Omega} v_r (R_1, \theta, \phi ) Y_{lm} ( \theta, \phi) d\Omega,
\eeq
where $v_r (R_1, \theta, \phi )$ is the radial velocity of fluid relative to the donor's core as a function of $\theta$ and $\phi$ at the donor's original radius, $r=R_1$. 
The power spectrum is computed as 
\beq\label{Sff}
S_{ff}(l) = \sum_{m=-l}^{l} v_{lm}^2,
\eeq
which implies a normalization such that the sum of $S_{ff}$ over $l$ is equal to the integral of the square of $v_r$ divided by its angular area, $\int v_r^2 / 4\pi$. See  \citet{shtools} for full description of the implementation.

To further aid in the interpretation of excited oscillations, we need to know eigenfrequencies of the donor star. In our case, the relevant frequencies are those of an artificial star with core removed -- as adopted in our gas dynamical model. We convert our stellar model to FGONG file format, and use the stellar-oscillation code GYRE to solve for the structure's eigenfrequencies \citep{2013MNRAS.435.3406T,2018MNRAS.475..879T}.\footnote{online at: https://bitbucket.org/rhdtownsend/gyre}  We adopt a vacuum outer boundary and zero radial displacement inner boundary, located at $0.3R_1$, to represent the simulated system. We further solve for mode frequencies under the Cowling approximation, which entails neglecting the oscillations perturbation on the star's gravity and is equivalent to the static self-gravity applied in our hydrodynamic model.  

In general, stellar oscillatory modes may be restored by pressure ($p$-modes), gravity ($g$-modes), or both (fundamental or $f$-modes).  Only pressure modes are supported within an isentropic star.  The fundamental (pressure) modes lack radial nodes, and we observe that these modes are most strongly excited in our simulated systems.

\section{Resonant Tidal Excitation During Binary Coalescence}\label{sec:resonances}

Here we discuss the conditions for resonance between the tidal forcing field and oscillation modes in the decaying binary. We demonstrate that such a binary crosses resonance with a number of high azimuthal-order modes and trace these resonantly-excited modes in our hydrodynamic model binary system. 

\subsection{Resonance Condition}

\begin{table}[tbp]
\begin{center}
\begin{tabular}{ccccc}
\hline
$l$ & $\omega_\alpha$ & $Q_\alpha$ & $M_\alpha$ & $C_\alpha$ \\
\hline
2 & 1.918 & 0.273 & $3.39 \times 10^{-3}$  & 0.4575 \\
3 & 2.197 & 0.259 & $1.86 \times 10^{-3}$  & 0.3117 \\
4 & 2.426 & 0.247 & $1.20 \times 10^{-3}$  & 0.2379 \\
5 & 2.630 & 0.237 & $8.36 \times 10^{-4}$  & 0.1926 \\
6 & 2.818 & 0.229 & $6.14 \times 10^{-4}$  & 0.1618 \\
\hline
\end{tabular}
\end{center}
\caption{Properties of $l=m$ fundamental modes of our non-rotating stellar model calculated with GYRE, in units of $G=M_1=R_1=1$.  Here $\omega_\alpha$ is the mode frequency, $Q_\alpha$ is the overlap integral between the tidal potential and the mode (equation \ref{eq:defQ}) ,  and $M_\alpha$  is the mode mass (equation \ref{eq:defmodeMass}). The coefficient $C_\alpha$ pertains to the perturbation of mode frequencies by rotation, and is described by equation \eqref{eq:Cnl}.  }
\label{table:modeproperties}
\end{table}

The characteristics of fundamental oscillation modes of different degree, $l$, and azimuthal order $m=\pm l$, within our stellar model are given in Table \ref{table:modeproperties}.   The mode frequencies, $\omega_\alpha$, depend only on $l$ (and not $m$) because our model is non-rotating (In general, rotation introduces frequency splittings for modes of different $m$). Because of the equatorial symmetry of the binary gravitational field, excited modes tend to have order of $m=\pm l$, therefore we focus on those modes here. In addition to mode frequency, we also tabulate the overlap between the tidal potential and the mode eigenfunction, $Q_\alpha$, equation \eqref{eq:defQ},  the mode mass, $M_\alpha$, equation \eqref{eq:defmodeMass}, and $C_\alpha$, which describes the effect of rotation on mode frequencies, equation \eqref{eq:Cnl}. These properties will be used later to compare the simulated result to predictions from linear theory. 

The force driving waves within the stellar structure is the binary's gravitational field. The frequency at which this forcing is applied is related to the binary's orbital frequency, approximately given by
$\Omega_{\rm orb} \approx \sqrt{GM/a^3}$
where $M=M_1+M_2$ is the total binary mass and $a$ is the separation. 
Resonances occur when modes remain in phase with the forcing \citep[first introduced by][in the context of non-radial stellar oscillations]{1941MNRAS.101..367C}, i.e. when 
\beq\label{resonance}
m \Omega_{\rm orb}  = \omega_\alpha,
\eeq
for a non-rotating star. 
This expression implies that the angular velocity of the forcing as viewed in the frame of the non-rotating stellar fluid, $\Omega_{\rm orb}$, is the same as the mode pattern frequency, $\omega_\alpha / m$.

\begin{figure*}[tbp]
\begin{center}
\includegraphics[width=0.45\textwidth]{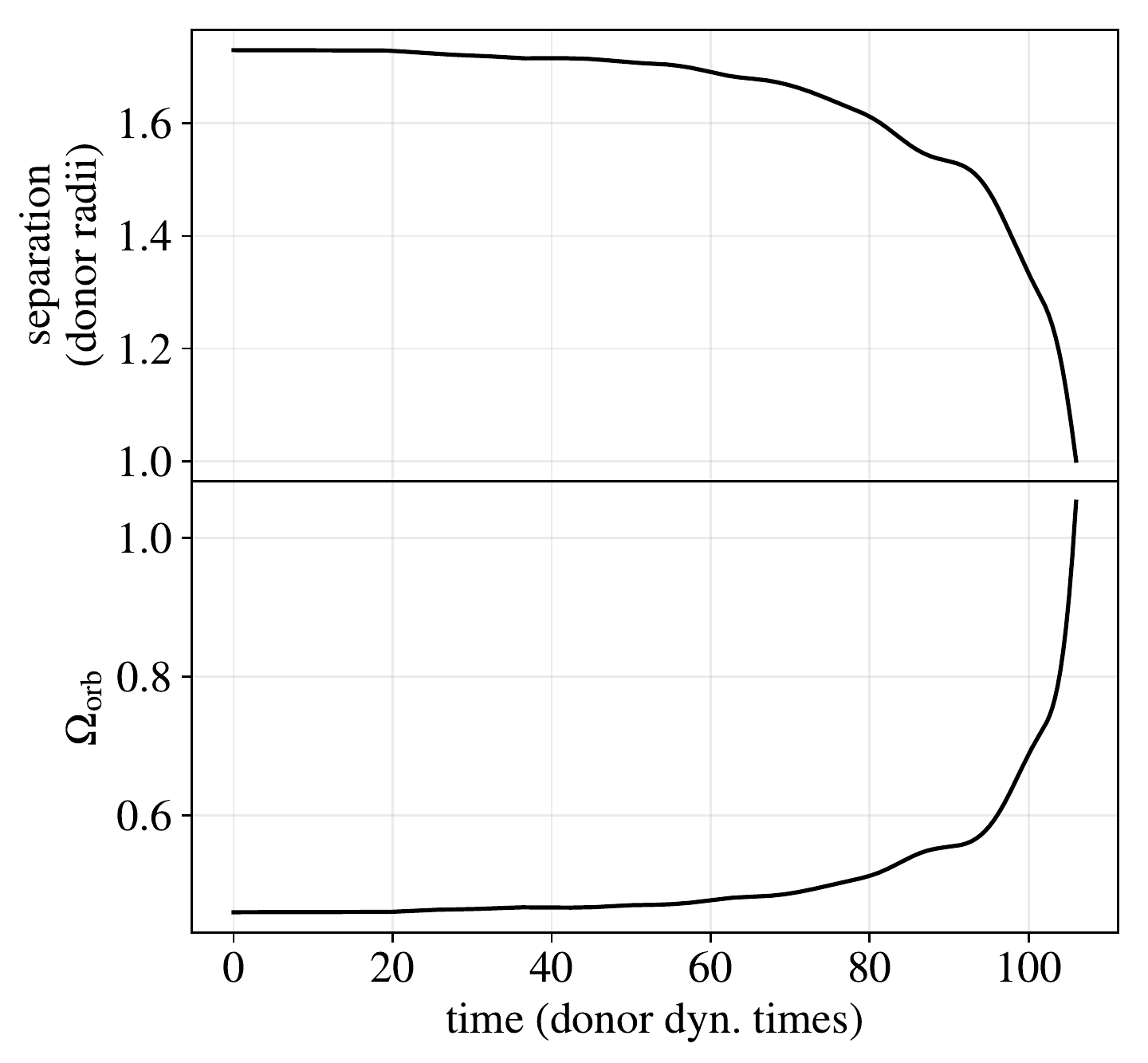}
\hspace{0.5cm}
\includegraphics[width=0.51\textwidth]{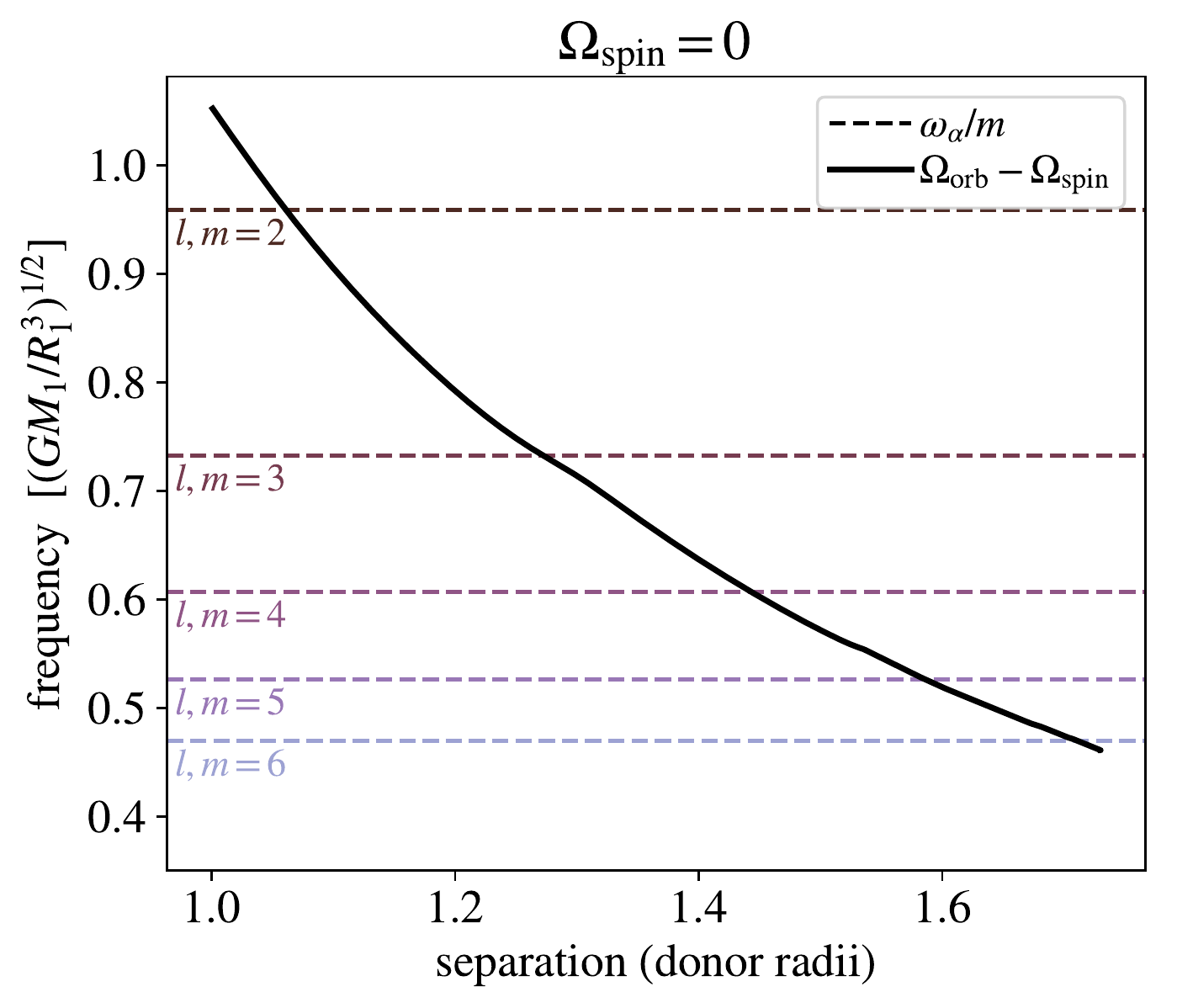}
\caption{Orbital properties and resonance crossing in a coalescing binary system.  The left panels show orbital separation and orbital frequency for our model binary system, which is undergoing runaway mass and angular momentum exchange that leads to its coalescence. The right panel compares the tidal forcing frequency, $\Omega_{\rm orb}-\Omega_{\rm spin}$, to mode frequencies $\omega_\alpha / m$ (here we examine the case $\Omega_{\rm spin}=0$). When the forcing and oscillation frequencies align, the system crosses through resonance with a particular mode, see equation \eqref{resonance}. High azimuthal order modes are in resonance at larger separations. As the binary orbit tightens, the system crosses through resonance with successively lower-order oscillation modes. }
\label{fig:resonancecrossing}
\end{center}
\end{figure*}

\subsection{Orbital Decay and Resonance Crossing}

Mass removal from the donor star leads to runaway orbital decay in  our simulated binary system \citep{2018ApJ...863....5M,2018arXiv180805950M}. As the binary orbit tightens, the donor increasingly overflows its Roche lobe, leading to accelerated mass loss and orbital decay. As the orbit tightens, the orbital frequency changes, roughly as $\Omega_{\rm orb} \propto a^{-3/2}$ \citep[despite some mass loss and the extended donor star structure; see Figure 16 of][]{2018ApJ...863....5M}.  The constantly changing orbital frequency implies that the forcing frequency of waves within the donor's envelope is always evolving and may, therefore, cross through resonances with particular modes. 

Figure \ref{fig:resonancecrossing} examines the decaying binary orbit in our simulation and its implications for mode resonance crossing. The left panels show binary separation and orbital frequency as a function of time. The simulation begins at approximately the analytic Roche limit of 1.73 donor radii \citep{1983ApJ...268..368E}. The binary separation then decays slowly, initially over many tens of donor dynamical times. Eventually, the orbital decay rate increases dramatically, and the accretor plunges within the envelope of the donor \citep{2018ApJ...863....5M}. 

In the right-hand panel of Figure \ref{fig:resonancecrossing}, we examine the terms of the resonance condition, equation \eqref{resonance}. We plot the forcing frequency, $\Omega_{\rm orb} - \Omega_{\rm spin}$, equivalent to $\Omega_{\rm orb}$ in this restricted case where $\Omega_{\rm spin}=0$. With horizontal lines we plot the eigenfrequencies of modes divided by their azimuthal order, $\omega_\alpha/m$. 

As the binary separation shrinks, the forcing frequency crosses resonances with different modes, beginning with higher azimuthal order modes at larger separations and progressing toward  low-order modes at small separations. At the initial separation of 1.73 donor radii, the system is near resonance with the $l,m=6$ mode. By the time the binary separation shrinks to just outside the donor's original radius, the nearest resonance crossing is with the $l,m=2$ mode.

\subsection{Mode Excitation During Coalescence}

\begin{figure*}[tbp]
\begin{center}
\includegraphics[width=0.49\textwidth]{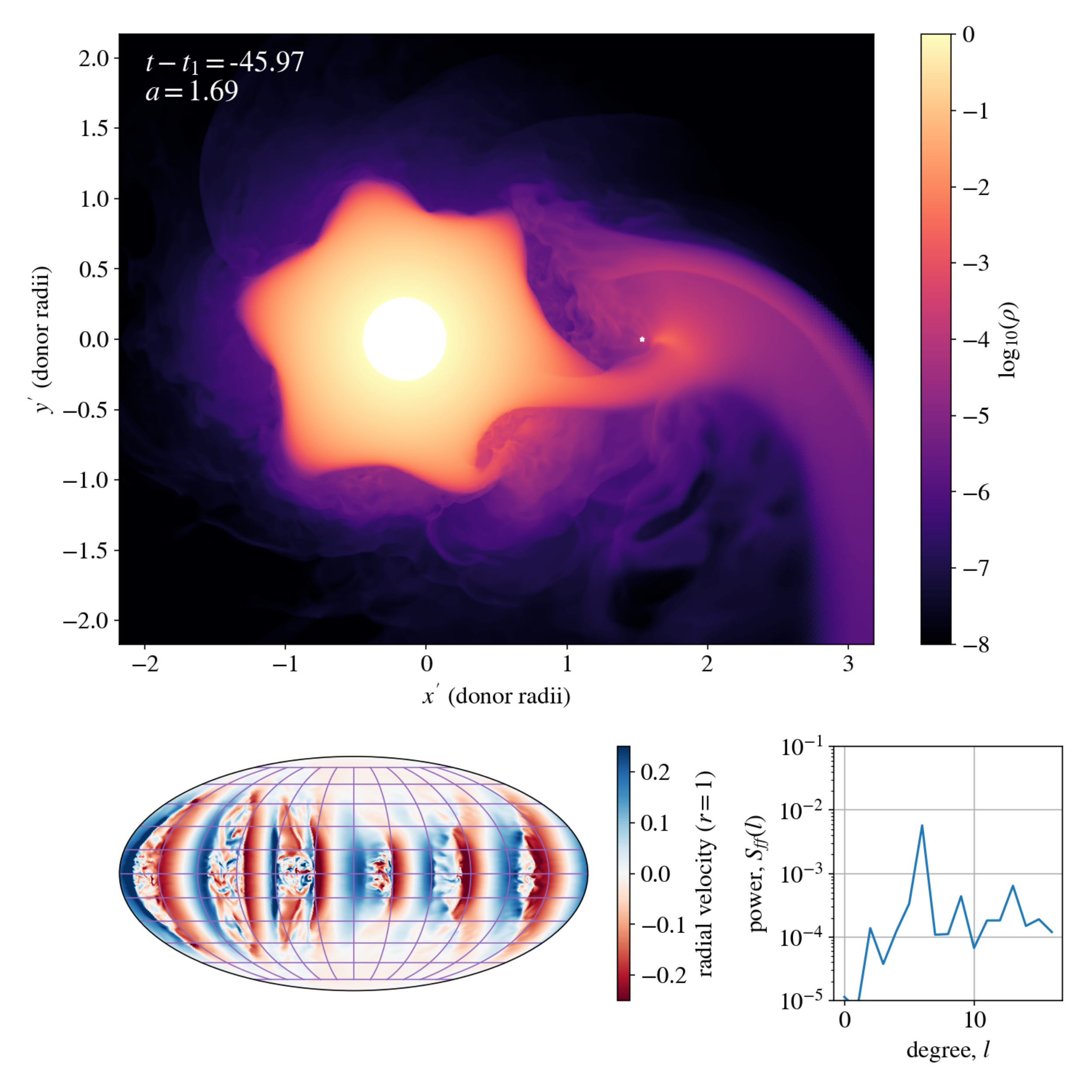}
\includegraphics[width=0.49\textwidth]{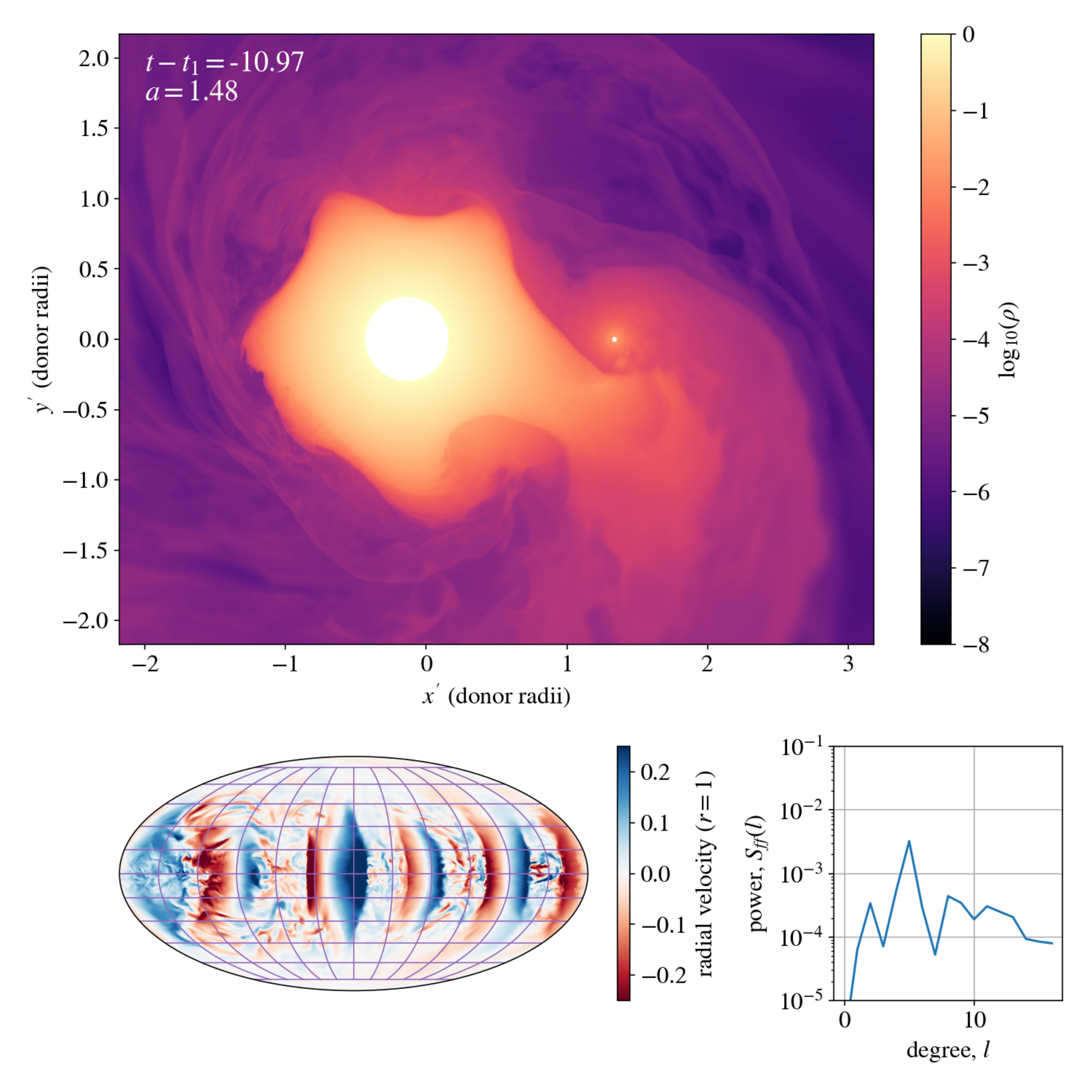}
\includegraphics[width=0.49\textwidth]{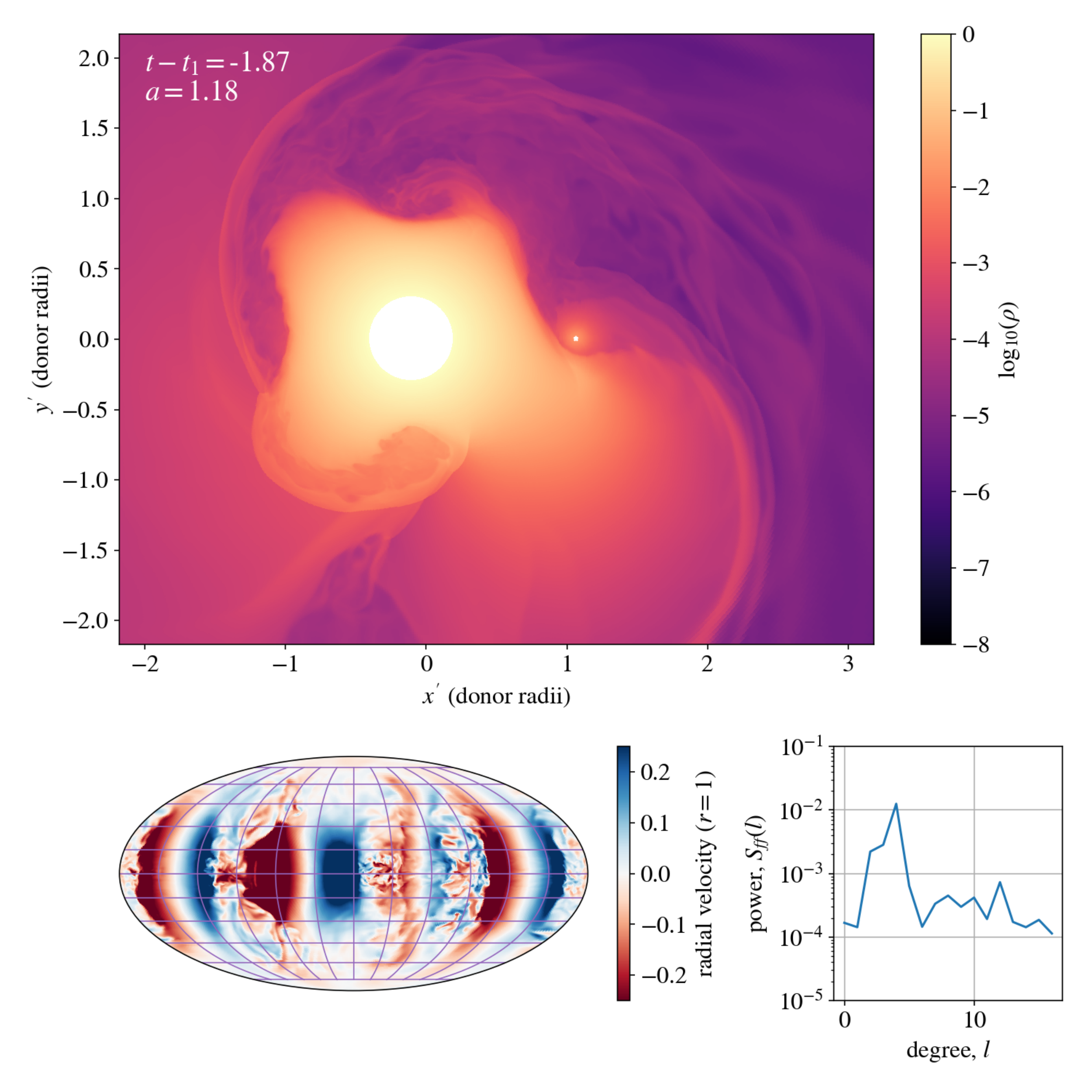}
\includegraphics[width=0.49\textwidth]{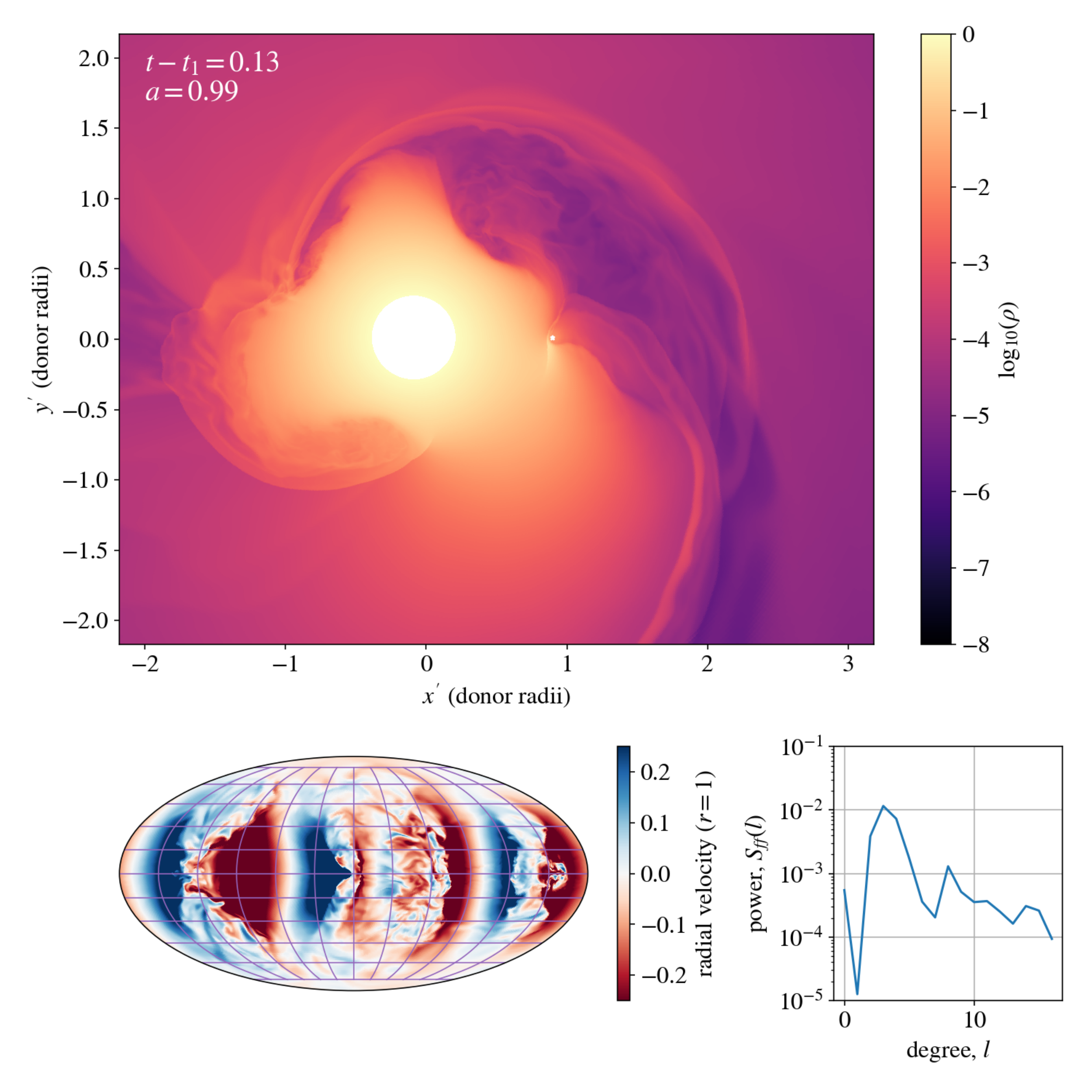}
\caption{Slices through the simulation orbital plane at increasing time (and decreasing binary separation). The large panels in each time show density in a slice through the binary equator. The lower panels show radial velocity on a spherical surface, $r=R_1$, and the power spectrum of this velocity field, $S_{ff}(l)$, as defined by equation \eqref{Sff}.   Large amplitude modes of different azimuthal order are clearly excited in these snapshots. As the binary separation decreases and the system sweeps through various resonances, the most excited mode transitions from higher to lower azimuthal order.    }
\label{fig:slice}
\end{center}
\end{figure*}

Figure \ref{fig:slice} shows a time series of properties of envelope of the donor star in the binary system. The large panels show gas density in a slice through the orbital midplane. The lower panels show radial velocity on a spherical surface (lower left) at $r=R_1$ and the spherical harmonic power spectrum, $S_{ff}(l)$, equation \eqref{Sff}, of this velocity field (lower right). 

The gas density slices of Figure \ref{fig:slice} reveal strongly-excited waves within the donor-star's envelope. The wave nearest the accretor can be seen to break and overflow toward the accretor. These waves can be traced to the $r=R_1$ radial velocity field by alternating regions of positive and negative radial velocity, measured relative to the donor's core in the orbiting system. The power spectrum plot reveals that much of the power in the velocity field is carried by particular degrees, $l$, of the spherical-harmonic decomposition. 

\begin{figure}[tbp]
\begin{center}
\includegraphics[width=0.49\textwidth]{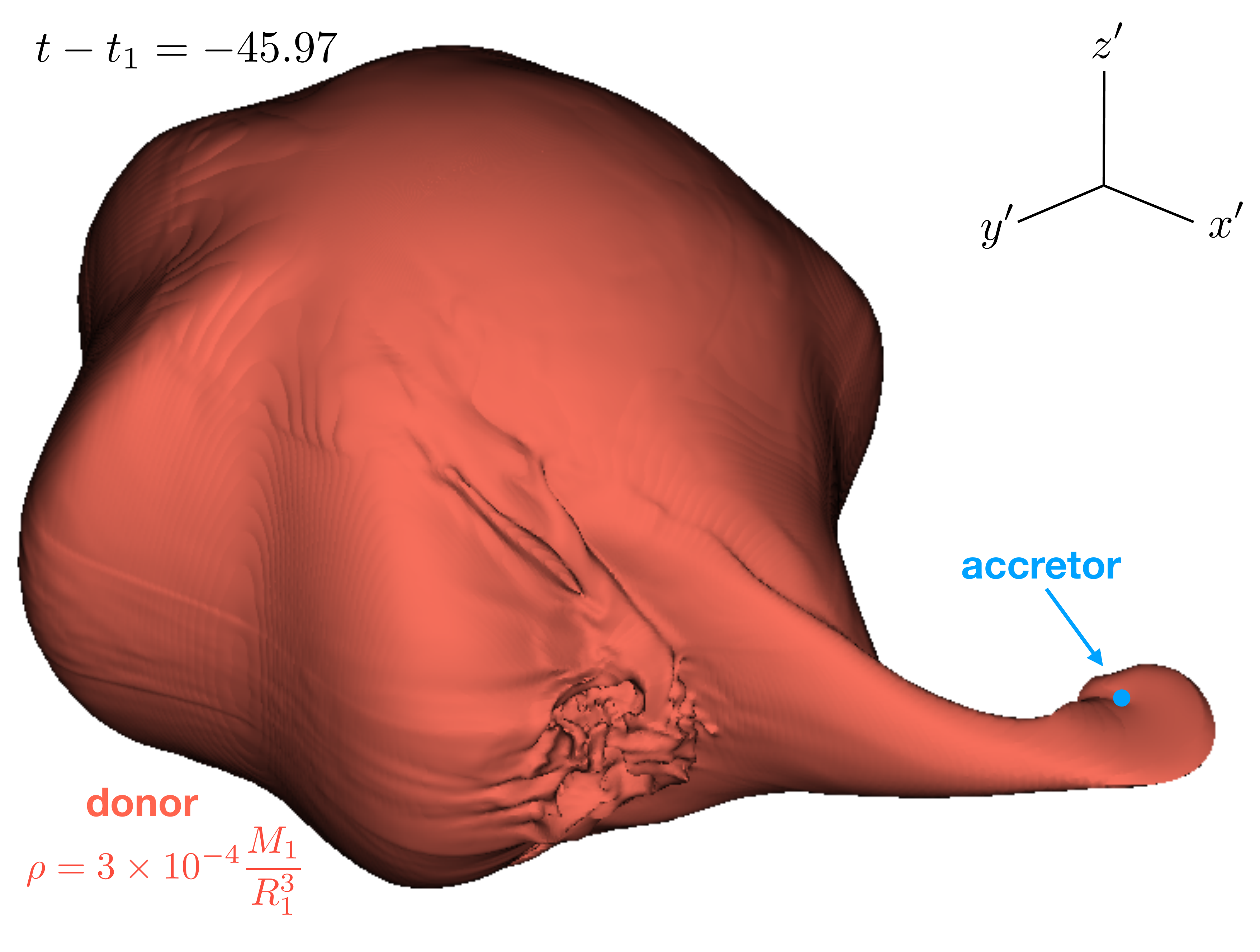}
\caption{Three-dimensional rendering of a density isosurface $\rho = 3\times 10^{-4}M_1/R_1^3$ for the same snapshot as shown in the upper left of Figure \ref{fig:slice}. This projection highlights the fact that mode amplitudes are largest at the equator of the donor star, and the breaking wave that trails behind the mass-transfer stream.  }
\label{fig:iso}
\end{center}
\end{figure}

Figure \ref{fig:iso} shows a three-dimensional rendering of a density isosurface of the first snapshot shown in Figure \ref{fig:slice}. The three-dimensional structure reveals that mode amplitudes are largest in the binary equatorial plane and vanish near the donor star's pole. Figure \ref{fig:iso} also highlights the three-dimensional structure of the mass-transfer stream, which is tidally compressed as it approaches the accretor, and the subsequent, breaking wave that trails the stream.

A comparison of different panels of Figure \ref{fig:slice} reveals that the number of peaks of the excited waves changes as the binary coalesces. While in the upper right panel there are six wave peaks ($l,m=6$), as the binary comes together we trace the appearance of five, then four, then three, giving the equatorial structure of the star a polygram shape. Waves also increase in amplitude  as the binary separation shrinks: both in height relative to the donor radius, and in magnitude of peak velocity.  Turning to the power spectrum for a more quantitative description, we see that the spherical-harmonic degree with the highest power also shifts, starting with higher orders and moving lower as the binary separation decreases.

\begin{figure}[tbp]
\begin{center}
\includegraphics[width=0.49\textwidth]{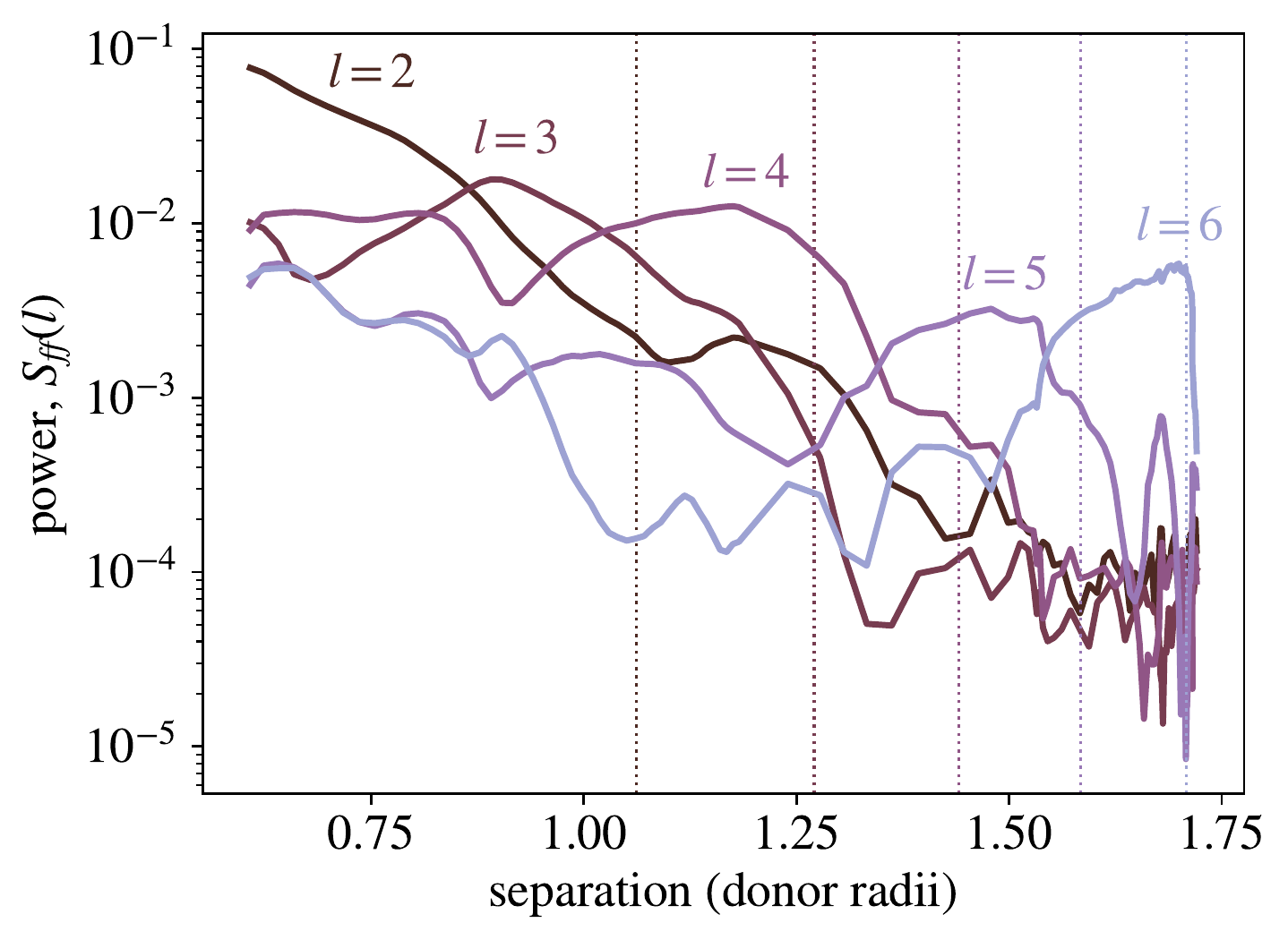}
\caption{Power, $S_{ff}(l)$, equation \eqref{Sff}, in the $r=R_1$ radial velocity field for spherical harmonic degrees from 2 to 6 as a function of binary separation. Initially (at large separation) most of the oscillatory motion is in the $l=6$ mode. As the binary separation decreases, successively lower $l=m$ modes become dominant. The separation of peak power in a mode lies inside the resonance crossing separations, marked with vertical dashed lines. Though modes are initially excited near the resonance crossing, continued orbital decay implies that by the time a mode pattern is fully established, the orbit is at smaller separation than the resonance crossing.  }
\label{fig:powersep}
\end{center}
\end{figure}

Figure \ref{fig:powersep} highlights the decomposition of the $r=R_1$ velocity field into its spherical harmonic components. We observe a shift in $S_{ff}(l)$ a function of binary separation. The most excited mode transitions as the binary separation decreases, such that the most-excited mode transitions from $l=6$ to $l=2$ by the time that the accretor is engulfed within the donor's envelope. 

Resonance crossings are indicated by vertical dashed lines in Figure \ref{fig:powersep} (the binary separations that satisfy the resonance condition of equation \eqref{resonance}). Interestingly, peak power within a given spherical harmonic order does not tend to occur at the separation of resonance crossing, but instead lies at somewhat smaller separation. For the lower-order modes this offset is particularly pronounced.  Although modes are resonantly excited principally at the resonance crossing locations, the mode pattern itself takes time to set up and propagate around the donor star.

\section{Mode Amplitudes in Decaying Binary System}\label{sec:modes}

Resonance crossing excites fundamental oscillation modes of a range of azimuthal orders during the binary's trend toward merger. We now contextualize these findings from our simulation model by comparison with the predictions of linear theory. 

\subsection{Predictions from Linear Theory}\label{sec:linear}
	We can predict the amount of energy transfer to a mode during a resonance crossing using linear hydrodynamics. We follow a formalism introduced by \citet{1994MNRAS.270..611L}. This general approach has since been applied to numerous problems involving white dwarf binaries \citep{2011MNRAS.412.1331F,2012MNRAS.421..426F,2012ApJ...756L..17F,2013MNRAS.430..274F,2013MNRAS.433..332B,2014MNRAS.444.3488F,2017MNRAS.468.2296V}, 
neutron star binaries \citep{1994ApJ...426..688R,1999MNRAS.308..153H,2006PhRvD..74b4007L,2013ApJ...769..121W,2017MNRAS.464.2622Y,2017MNRAS.470..350Y,2017PhRvD..96h3005X}, 
stellar binaries \citep{1997ApJ...490..847L,2012MNRAS.420.3126F,2017MNRAS.472.1538F,2018MNRAS.473.5165H, 2018MNRAS.476..482V,2018MNRAS.481.4077S}, 
and planet-star systems \citep{2004ApJ...610..477O,2012MNRAS.423..486L,2012ApJ...751..136W,2016ApJ...816...18E,2017ApJ...849L..11W,2018AJ....155..118W, 2018MNRAS.476..482V,2018arXiv181205618V}.  
	
	 For comparison with the numerical model, we take $\Omega_{\rm spin}=0$ in the following expressions. The gravitational potential produced by $M_2$ can be expanded in a spherical harmonic basis as 
	\begin{equation}
	U(\boldsymbol{r},t) = - GM_2\sum_{ml} \frac{W_{lm} r^l}{D(t)^{(l+1)}} \text{e}^{- \text{i} m \Phi(t)} Y_{lm} (\theta,\phi), \label{eq:potential}
	\end{equation}
	where $\textbf{r} = (r,\theta,\phi)$ is the position vector in spherical coordinates with respect to the center of mass of the donor star, $M_1$, $\Phi(t) = \int dt\; \Omega_{\rm orb }t$ is the orbital true anomaly and 
	\begin{align}
	W_{lm} =& (-1)^{(l+m)/2}\left[\frac{4\uppi}{2l+1}(l+m)!(l-m)!\right]^{1/2} \nonumber \\ & \times \left[2^l\left(\frac{l+m}{2}\right)!\left(\frac{l-m}{2}\right)!\right]^{-1}.
	\end{align}
	
The perturbation to the donor-star's fluid is specified by the Lagrangian displacement vector $\boldsymbol{\xi}(\boldsymbol{r},t)$. We can decompose $\boldsymbol{\xi}(\boldsymbol{r},t)$ into normal modes $\boldsymbol{\xi}_\alpha(\boldsymbol{r} , t) \propto \text{e}^{\text{i} m \phi-\text{i}\omega_\alpha t} $ of frequencies $\omega_\alpha$, where $\alpha = \{n_rlm\}$ is the standard mode index. We adopt the convention that a mode with $\omega_\alpha>0, m>0$ is prograde. 
\citet{2006PhRvD..74b4007L} derived the most general expression for the energy transfer to a mode $\alpha$ when the orbit sweeps through the resonance $\omega_\alpha = m \Omega_{\rm orb}$ (see their Section II; for a non-rotating star, this expression reduces that in \citet{1994MNRAS.270..611L}). 

The predicted energy transferred to a mode is proportional to the time spent near resonance. The resonance crossing time for the $l=m$ mode is
\beq\label{tres}
\Delta t_l = \left ( \frac{2 P_{\rm orb}}{3l} \frac{a}{|\dot a| }  \right)^{1/2}.
\eeq
It is therefore related to the geometric mean of the orbital period and the orbital decay rate, $a / |\dot a|$ \citep{2006PhRvD..74b4007L}, where all quantities are evaluated at the resonance separation. One underlying assumption of the linear theory relates to the relative timescale of resonance crossing, $P_{\rm orb} \ll \Delta t_l \ll a / |\dot a|$. 
To leading order in the binary separation $a$, the energy transfer to the  $l=m$ mode is given by 
	\begin{align}
		\Delta E_{\rm \alpha} =& \frac{GM_2^2}{R_1} \left(\frac{4\uppi^2}{3l}\right)\left(\frac{M_1}{M}\right)\nonumber \\
	&\times \left|\frac{a}{\dot{a}P_{\rm orb}}\right|_l W_{ll}^2 Q_\alpha^2 \left(\frac{R_1}{a_{l}}\right)^{2l-1}, \label{eq:DeltaE_alpha}
	\end{align}
	where $a/(\dot{a} P_{\rm orb})$ is evaluated at the resonance separation $a_l = (GM)^{1/3}(\omega_\alpha/l)^{-2/3}$ (see equation \eqref{resonance} for $l=m$), and 
	\begin{equation} \label{eq:defQ}
	Q_\alpha \equiv \int d^3x \; \rho \boldsymbol{\xi}_\alpha^*\cdot\nabla(r^lY_{lm}),
	\end{equation}
which is tabulated in Table \ref{table:modeproperties}.	In equation (\ref{eq:defQ}), $\boldsymbol{\xi}_\alpha$ is normalized such that $\int d^3x \; \rho \boldsymbol{\xi}_\alpha^*\cdot \boldsymbol{\xi}_\alpha = 1$, and $Q_\alpha$ is in units $G=M_1=R_1=1$ (identical to the hydrodynamic model code units).
	
	The amount of energy transferred to the mode dictates the size of the surface velocity perturbation $\delta \boldsymbol{v}(R1,\theta,\phi)$. For comparison with simulations, we are interested in the radial component of the velocity $\delta v_r(R_1,\theta,\phi) = \delta v_r(R_1)Y_{lm}(\theta,\phi)$. To relate the energy transfer and radial velocity perturbation at $r=R_1$, we define the mode mass
	\begin{equation}
	M_\alpha = \int d^3 x \; \rho \left|\frac{\boldsymbol{\xi}_\alpha}{\xi_{r}(R_1)}\right|^2, \label{eq:defmodeMass}
	\end{equation}
	where $\xi_{r}(R_1)$ is the radial component of the Lagrangian displacement vector at the surface. The mode masses for fundamental modes of our model star are tabulated in Table \ref{table:modeproperties}. Note that $|\delta v_r (R_1)|= \omega_\alpha |\xi_{r}(R_1)|$, and we have
	\begin{equation}
	|\delta v_r(R_1)| = \left(\frac{\Delta E_\alpha}{M_\alpha}\right)^{1/2}.
	\end{equation}
	We can find a RMS velocity at the stellar equator by fixing $\theta=\uppi/2$ and computing 
	\begin{align}\label{vrms}
	\delta v_{\rm RMS}(R_1) =& \left[\frac{1}{2\uppi}\int d\phi \; |\delta v_r(R_1)|^2|Y_{lm}(\uppi/2,\phi)|^2\right]^{1/2} \nonumber\\
	=& \frac{1}{2^l l!}\sqrt{\frac{(2l+1)!}{4\uppi}} \left(\frac{\Delta E_\alpha}{M_\alpha}\right)^{1/2}.
	\end{align}
These expressions contain a complete prediction of the mode velocities at the stellar surface given a mode, $\alpha$, and knowledge of the mode profile and its properties.

\subsection{Comparison to Model System}

In this section, we employ the result of Section \ref{sec:linear} with input from our hydrodynamic model to directly compare the oscillatory surface velocities predicted by linear theory to those obtained in the simulation. 
As a first ingredient, Table \ref{table:modeproperties} displays the calculated mode properties $Q_\alpha$ and $M_\alpha$ for the stellar model described in Section \ref{sec:method} and employed in our hydrodynamic simulation. 

\subsubsection{Orbital Decay and Resonance Crossings}

In our model binary system, orbital decay arises primarily from mass removal from the donor by the gravitational force from the accretor. As discussed in Section, \ref{sec:resonances} this leads to runaway orbital tightening. This orbital decay allows the system to pass through resonances and  determines the total energy which may be transferred to a given mode.

\begin{figure}[tbp]
\begin{center}
\includegraphics[width=0.49\textwidth]{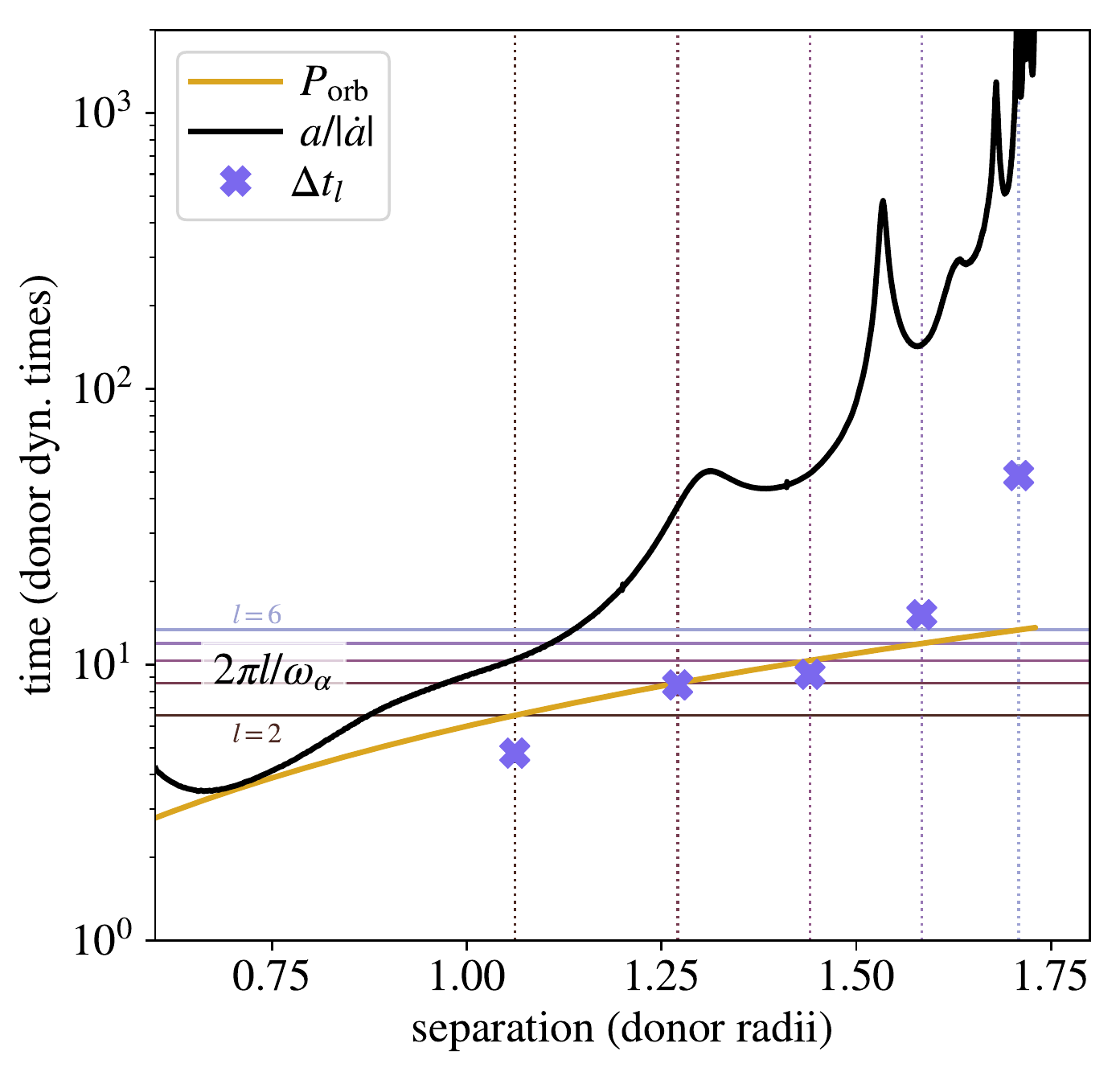}
\caption{Comparison of orbital decay timescale to orbital period as a function of binary separation. We also show the separations of resonance crossings (vertical lines) and $2\pi l/\omega_\alpha$, the time for an $l=m$ mode pattern to propagate azimuthally around the donor star.  The resonance crossing times, $\Delta t_l$, are marked as crosses, see equation \eqref{tres}. At large orbital separations, the orbital decay time is long compared to the orbital period. This implies that energy is deposited into oscillation modes over many orbits. On the other hand, at small separations the decay timescale approaches the orbital period and the mode propagation timescale, implying that mode patterns may be only partially established before the system drifts out of resonance.  }
\label{fig:tdecay}
\end{center}
\end{figure}

Figure \ref{fig:tdecay} examines how the ongoing orbital decay drives to mode excitation. The orbit first decays slowly at large separations (large  $a / |\dot a|$), then more quickly as the binary separation decreases.  Features of slower and more rapid orbital decay can be seen in the decay timescale of Figure \ref{fig:tdecay}, which will be discussed further in Section \ref{sec:modeorbit}. Figure \ref{fig:tdecay} also plots the binary orbital period and the resonance crossing time, equation \eqref{tres}.  We compare  these properties of the binary orbit to the timescale for a mode pattern to propagate around the whole star following excitation, $2\pi l / \omega_\alpha$.  Resonance crossing separations are indicated with vertical lines.  We note that at resonance crossing separations,  $2\pi l / \omega_\alpha = P_{\rm orb}$. 

At large separations, the resonance crossing timescale, $\Delta t_l$, is intermediate between the shorter orbital period and the longer orbital decay rate. Further, $\Delta t_l$  is longer than the mode pattern timescale (as indicated by horizontal lines of $2\pi l / \omega_\alpha$).  This hierarchy of timescales is in accordance with the assumptions that lead to equation \eqref{eq:DeltaE_alpha}. Because resonance crossing is long compared to the time that the mode takes to fully establish, we observe a fully developed mode pattern at larger separations in Figure \ref{fig:slice}. As the separation decreases, however, $\Delta t_l$ becomes similar to the orbital period. This implies a rapid sweep through resonance compared to the time for the mode to be fully excited about the star.

\subsection{Comparison of Predicted to Observed Oscillation Amplitudes}

\begin{figure}[tbp]
\begin{center}
\includegraphics[width=0.49\textwidth]{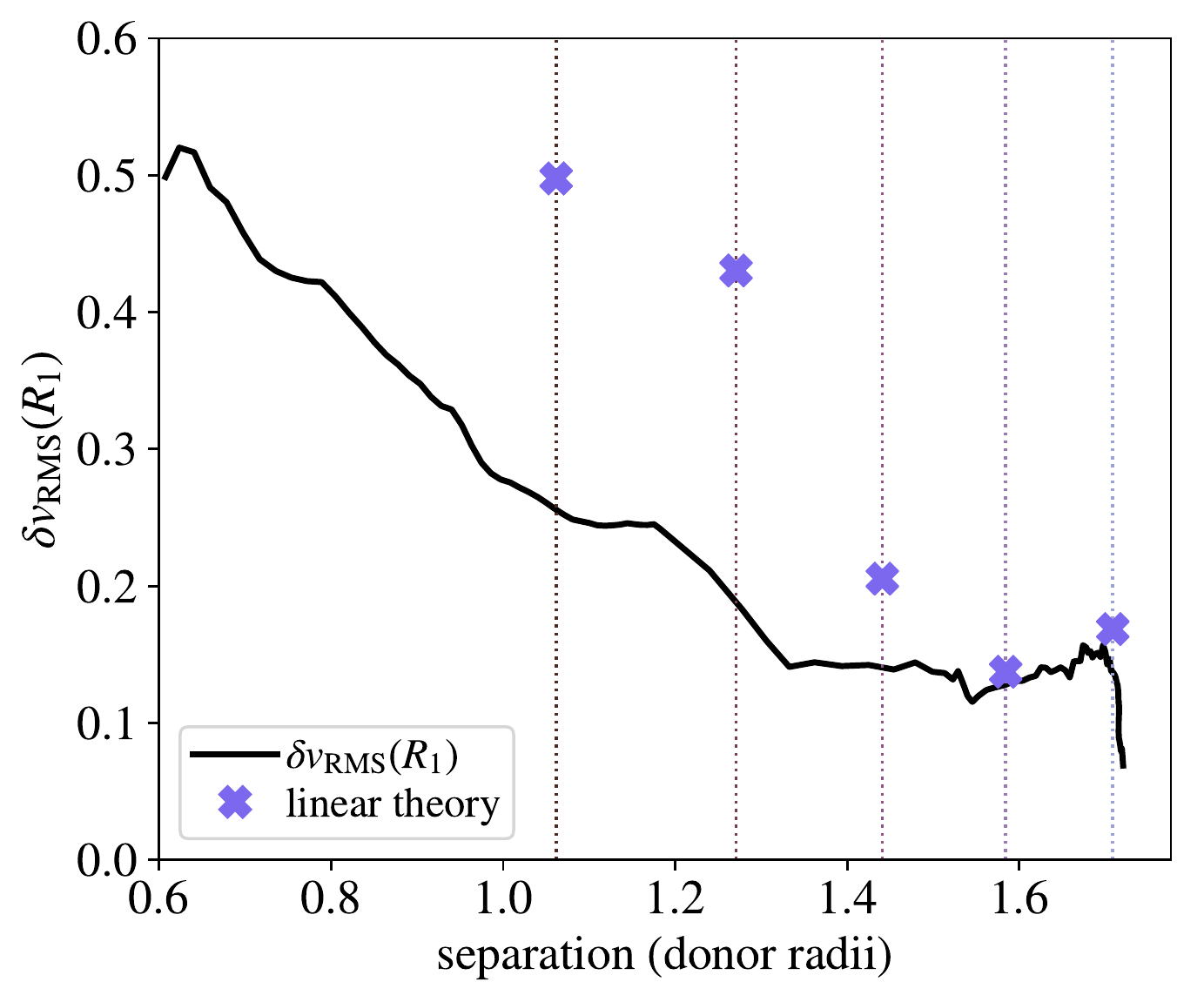}
\caption{Root-mean-square radial velocity in the donor star equator measured from the simulation and predicted from the linear theory, equations \eqref{eq:DeltaE_alpha} and \eqref{vrms}. The correspondence between the hydrodynamic model and the linear theory is striking; the linear theory reproduces the qualitative trend toward higher oscillatory amplitudes with decreasing separation and approximates $\delta v_{\rm RMS}(R_1)$ to within a factor of two for all modes. }
\label{fig:vrms}
\end{center}
\end{figure}

Together, these properties completely specify the input for the linear theory's predictions of the oscillatory velocity field. We use the RMS velocity on the stellar equator (in the plane of orbital motion, where the tides have maximum amplitude) as our point of comparison between hydrodynamic model and linear theory in Figure \ref{fig:vrms}. The equatorial RMS velocity is specified by equation \eqref{vrms} in the linear case. In the hydrodynamic model, we compute the RMS velocity of all zones in the shell with $r=R_1$ within $\pm \pi/32$ of the equator.

Figure \ref{fig:vrms}  shows that the linear theory captures both the qualitative trends and the approximate amplitude of the RMS velocity field. The prediction is particularly accurate for the $l,m=6$ and $l,m=5$ modes, while somewhat over-predicting the RMS velocity of the lower-order modes. This may relate to the comparison of the resonance crossing time to the orbital period. For the higher-order modes $P_{\rm orb} \lesssim \Delta t_l$ and the mode pattern is able to fully establish around the star. Orbital decay is so rapid through the lower-order resonance crossings that $P_{\rm orb} \gtrsim \Delta t_l$, and the mode pattern is only able to partially establish around the star's equator.  The lower values of simulated $\delta v_{\rm RMS}(R_1)$ may therefore reflect the partial establishment of modes given the rapid orbital decay. 
We note that while modes with higher azimuthal order than $m=2$ have been rarely considered in cases of tidal excitation, the linear model nonetheless encodes much of the key behavior when applied to  these higher-order modes.

\section{Discussion}\label{sec:discussion}

Here we explore some extensions of our basic findings, including the influence of a rotating donor star, coupling between oscillatory modes and the orbital evolution, dissipation of mode energies into the donor star gas, and the potential detectability of resonantly excited modes. 

\subsection{Influence of Donor-Star Rotation}\label{sec:rotation}

\begin{figure*}[tbp]
\begin{center}
\includegraphics[width=0.65\textwidth]{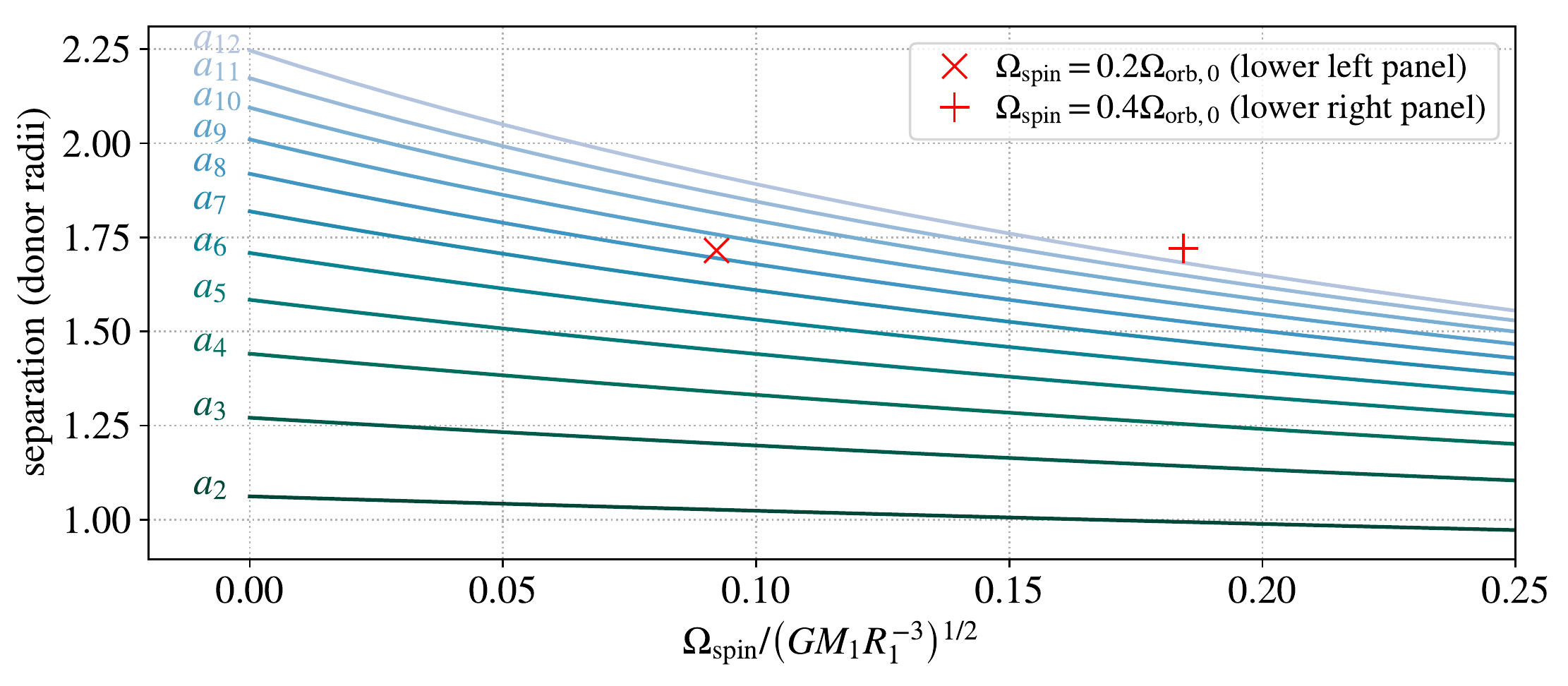}
\includegraphics[width=0.48\textwidth]{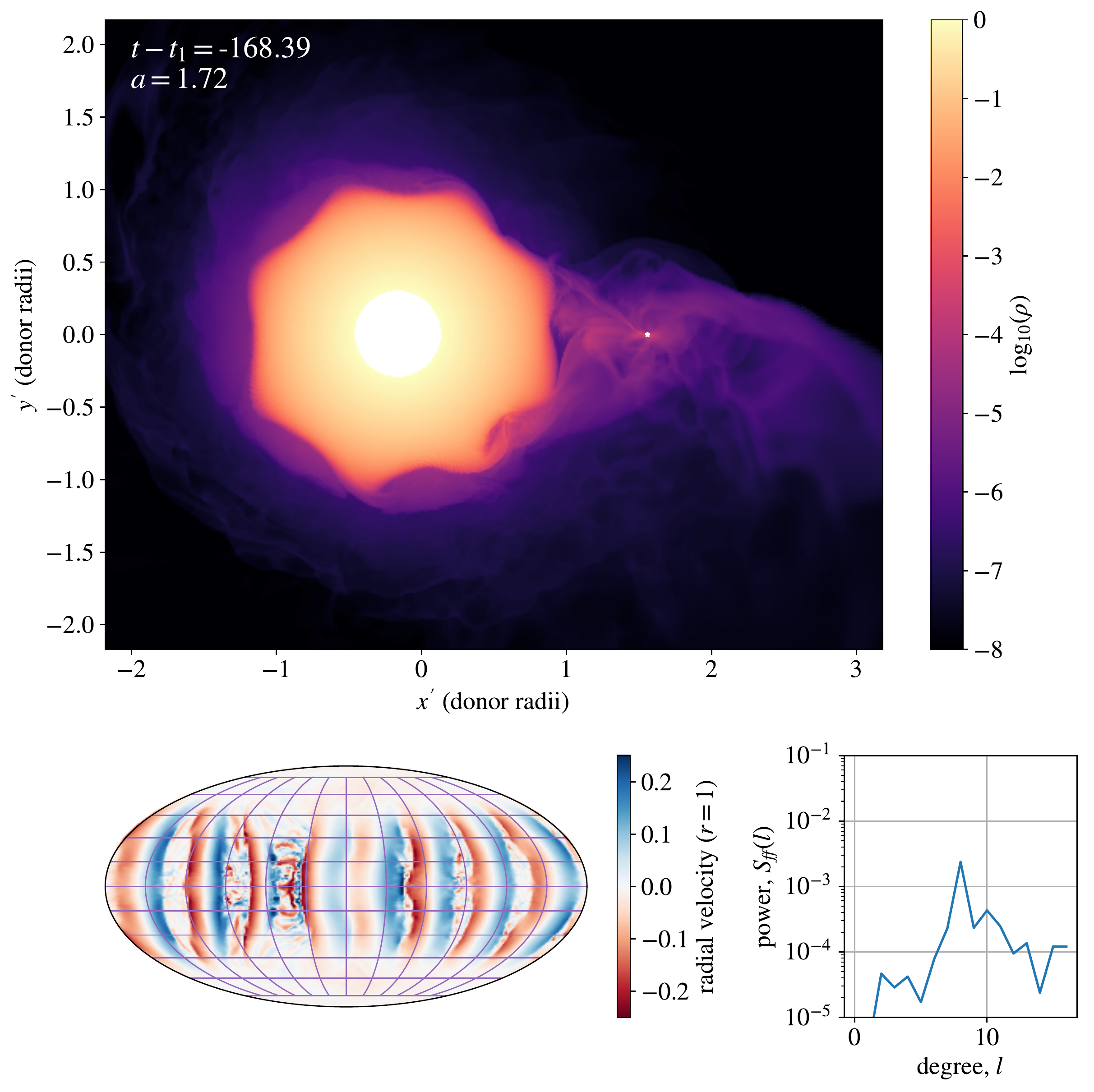}
\includegraphics[width=0.48\textwidth]{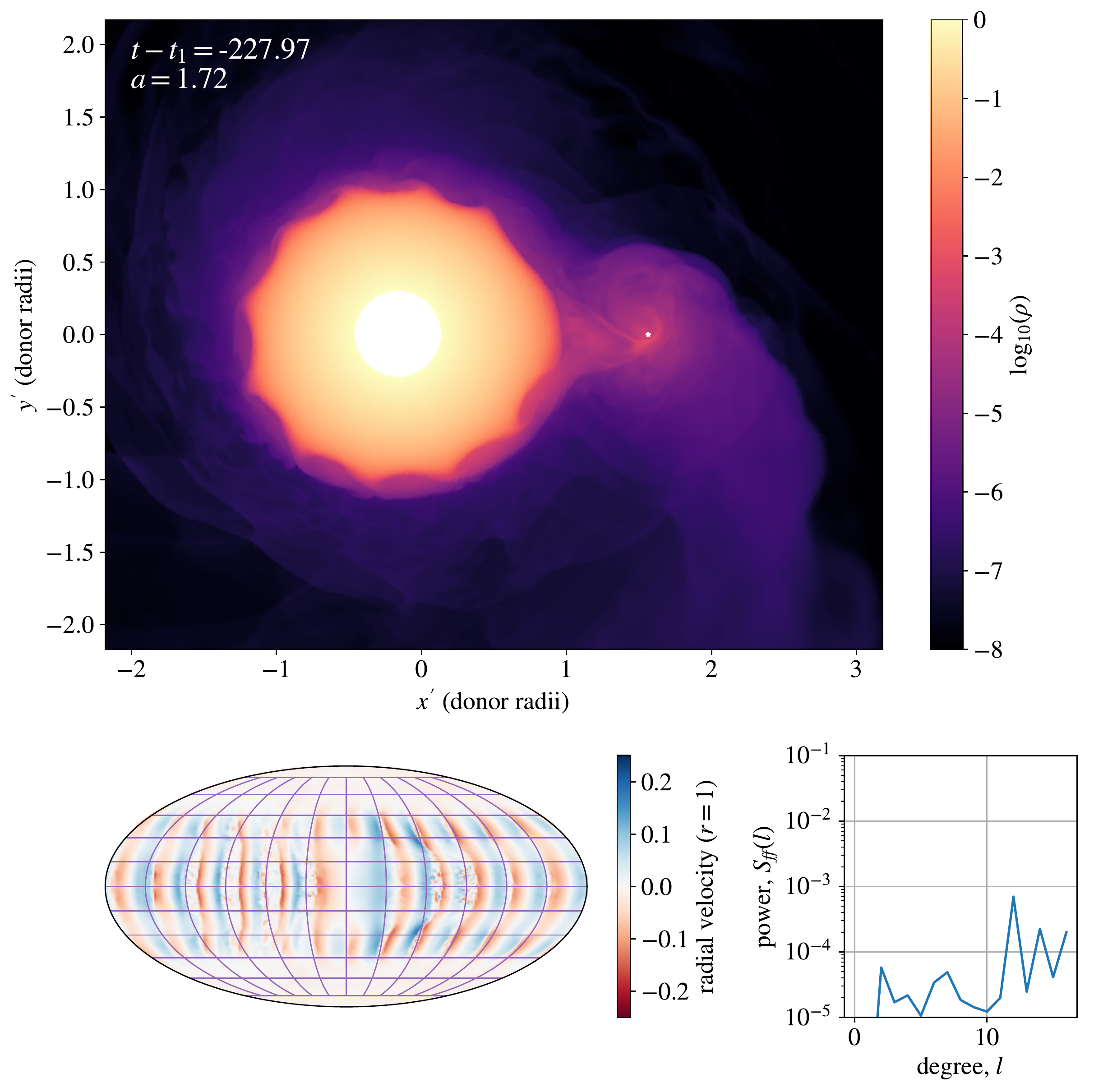}
\caption{Influence of donor rotation on resonantly-excited oscillations.  The upper panel shows how resonance crossings shift in separation with donor-star rotation. For donors that partially co-rotate with the orbital motion, higher-azimuthal order modes are excited at a given separation as rotational frequency increases. Two examples are shown in the lower panels: a donor star rotating at 20\% the initial orbital frequency ($\Omega_{\rm spin} =0.092 \left(GM_1R_1^{-3}\right)^{1/2}$; left panel) exhibits primarily a $l,m=8$ oscillation; a donor star rotating with 40\% the initial orbital frequency ($\Omega_{\rm spin} =0.184 \left(GM_1R_1^{-3}\right)^{1/2}$;  right panel) has a higher order wave excited with $l,m=12$.  These models may be compared to the non-spinning donor, which initially exhibits an $l,m=6$ oscillation.  }
\label{fig:rot}
\end{center}
\end{figure*}

For simplicity, we have so far considered a non-rotating donor star. However, the qualitative results of our study extend to the generalized scenario of a rotating donor with two key modifications \citep[e.g.][]{1997ApJ...490..847L}. First, the apparent forcing frequency in the frame of the stellar fluid changes. Secondly, rotation modifies the mode frequencies themselves. 

For a donor star in solid body rotation with frequency $\Omega_{\rm spin}$, the condition for resonance is 
\beq\label{resonancespin}
m \Omega_{\rm orb} = \sigma_\alpha,
\eeq
where $\sigma_\alpha$ is the mode frequency in the inertial frame. Equation \eqref{resonancespin} 
is equivalent to $m \left(\Omega_{\rm orb}-\Omega_{\rm spin}\right) = \omega_\alpha^{\rm rot}$, with $\omega_\alpha^{\rm rot}$ referring to the mode frequency in the rotating frame. 
 In the case of synchronized rotation,  $ \Omega_{\rm orb}= \Omega_{\rm spin}$,  the forcing has no apparent time variability for the stellar fluid.

Next we compute the rotational modification of the mode frequencies. When the unperturbed mode frequency, $\omega_\alpha^{(0)}$, is much greater than $\Omega_{\rm spin}$, we can treat the effect of rotation as a perturbation. The frequency in the inertial frame is then
	\begin{equation}
	\sigma_\alpha = \omega_\alpha^{(0)} + m \Omega_{\rm spin} (1-C_\alpha),
	\end{equation}
	where 
	\begin{equation}
	mC_\alpha\Omega_{\rm spin} = i\int d^3x \; \rho \boldsymbol{\xi}_\alpha^*\cdot(\boldsymbol{\Omega}_{\rm spin} \times \boldsymbol{\xi}_\alpha), \label{eq:Cnl}
	\end{equation}
	which is normalized such that  $\int d^3x \; \rho \boldsymbol{\xi}_\alpha^*\cdot \boldsymbol{\xi}_\alpha = 1$. The $C_\alpha$ for $l=m$ f-modes up to $l=6$ of the stellar model are shown in Table \ref{table:modeproperties}.

In Figure \ref{fig:rot}, we examine two example cases of an asynchronously rotating donor star. The upper panel shows the separation of $l=m$ fundamental mode resonance crossings as a function of donor spin, where $\Omega_{\rm spin}$ is oriented along the same axis as $\Omega_{\rm orb}$.  As the donor star rotates faster, the azimuthal order of the mode excited at a given separation increases. The lower panels show realizations of this prediction for a donor's spinning with 20\% and 40\% of the orbital frequency, respectively.  While in the non-spinning case, an $l,m=6$ oscillation was excited, when $\Omega_{\rm spin} = 0.2 \Omega_{\rm orb,0}$, an $l,m=8$ oscillation is excited shortly after the simulation is initialized. When $\Omega_{\rm spin} = 0.4 \Omega_{\rm orb,0}$, most of the oscillatory power is in the $l,m=12$ mode. The subsequent behavior of these models follows that of the non-spinning case, sweeping through decreasing-order resonances as the binary trends toward coalescence.

\subsection{Mode-Orbit Coupling}\label{sec:modeorbit}

When the orbit sweeps through a resonance, an amount of energy $\Delta E_\alpha$, equation \eqref{eq:DeltaE_alpha}, is transferred from the orbit to the mode, effectively coupling the orbital and oscillatory evolutions. Figure \ref{fig:tdecay} clearly shows peaks and troughs in the orbital decay rate, along with a general trend toward shorter decay time with decreasing binary separation. For example, the feature at separation between 1.6 and 1.5 donor radii represents a factor of approximately 5 slow-down in orbital decay rate. In what follows, we explore the origin of these features. 

\begin{figure}[tbp]
\begin{center}
\includegraphics[width=0.49\textwidth]{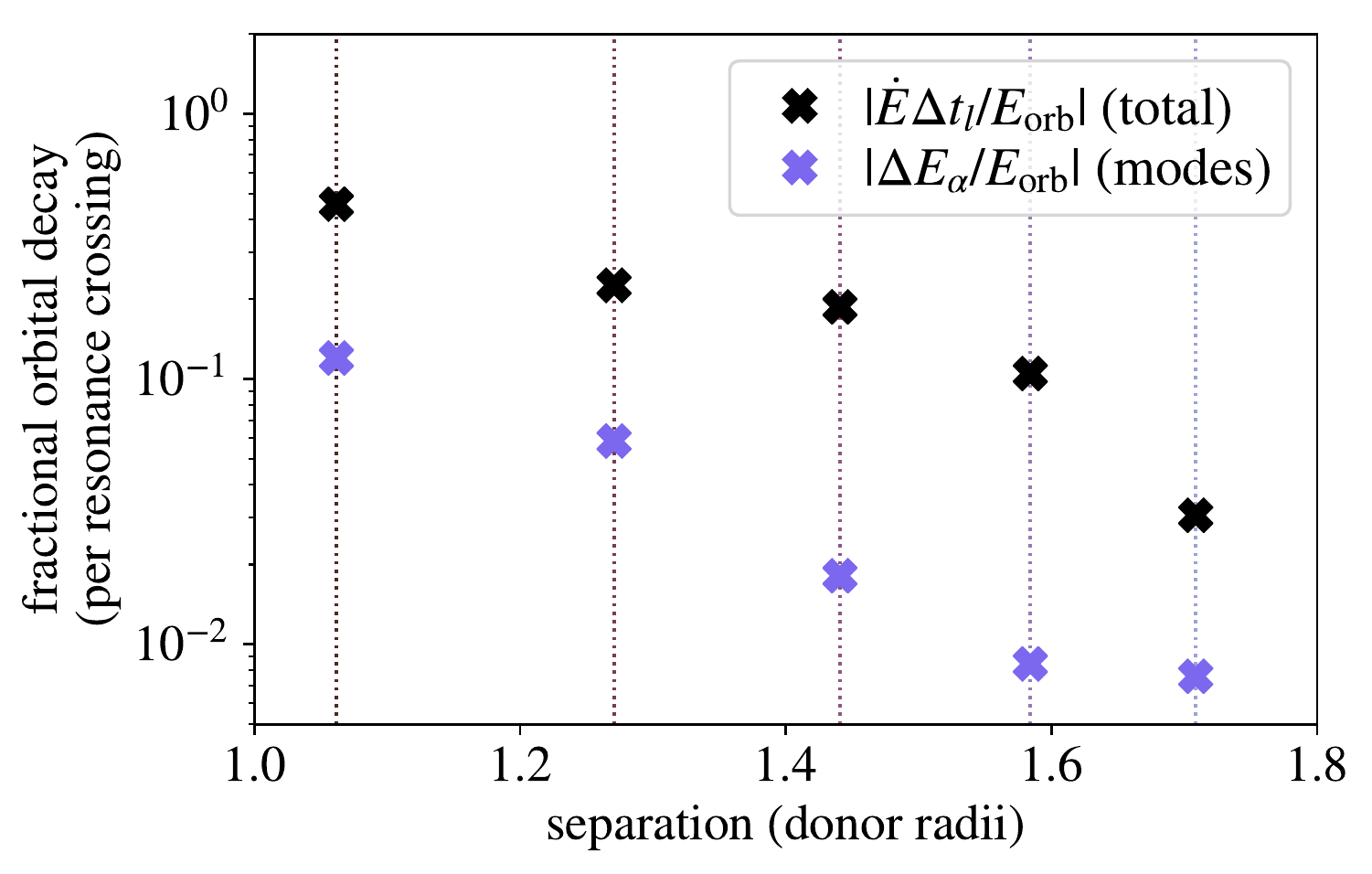}
\caption{Fractional orbital decay generated during resonance crossings. We compute the total orbital energy change during a given resonance crossing time, $\Delta t_l$ in our numerical model as $\dot E_{\rm orb} \Delta t_l $, and compare it to the deposition of energy into oscillatory motions, $\Delta E_\alpha$.  This comparison  shows that the energy deposited into modes accounts for only a small fraction of the orbital evolution. It therefore cannot explain the large-amplitude modulations of the orbital decay rate observed in Figure \ref{fig:tdecay}.  }
\label{fig:decaymode}
\end{center}
\end{figure}

Figure \ref{fig:decaymode} compares the total orbital energy change during resonance crossings to the energy transfer to the mode predicted by the linear model. The former is computed by multiplying the orbital decay rate, as shown in Figure \ref{fig:tdecay}, by the resonance crossing time, $\Delta t_l$. We observe that the total model orbital decay is always significantly larger than that which can be attributed to the modes. This conclusion implies that, in our situation, rapid non-conservative mass transfer instead drives the bulk of the orbital decay \citep[as discussed in][]{2018ApJ...863....5M,2018arXiv180805950M} and that that tidal exchange with oscillatory motions in the stellar envelope has negligible effect on the orbit.

\begin{figure}[tbp]
\begin{center}
\includegraphics[width=0.49\textwidth]{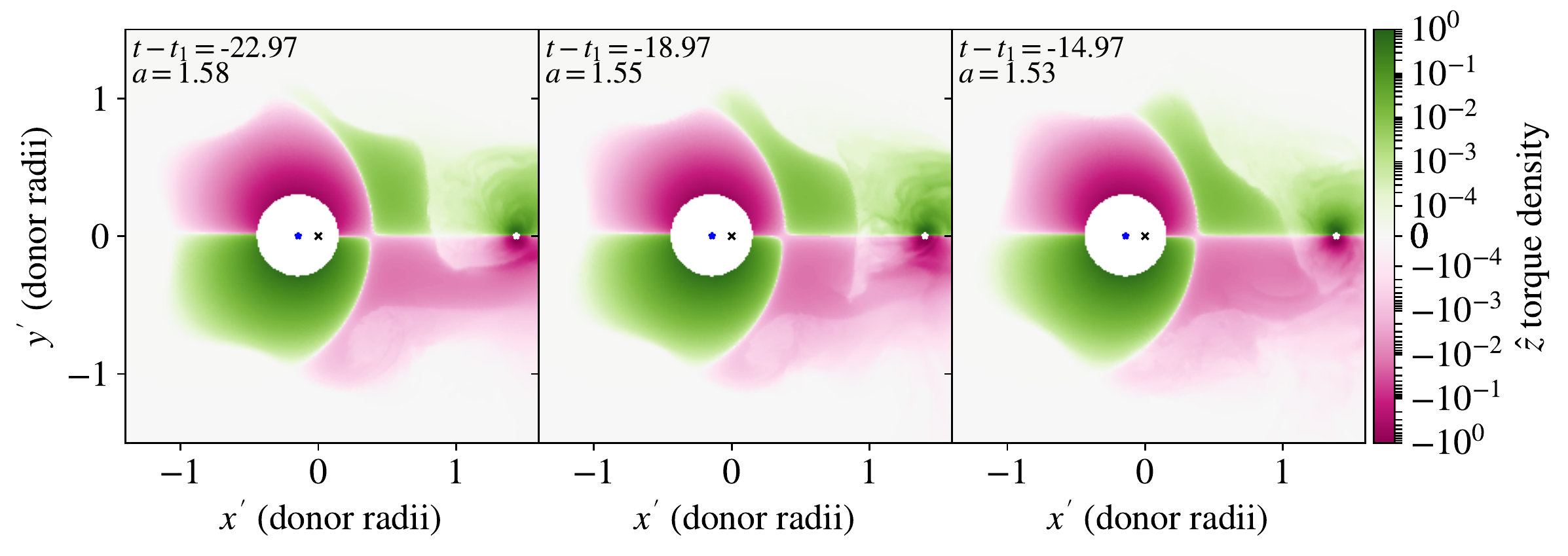}
\caption{Non-linear mode-orbit interactions. Slices show the local density of torquing material. Positively-torquing material pull the orbit to higher angular momenta, negatively-torquing material pulls the orbit to decay to lower angular momenta. The net balance drives the overall orbital evolution.
These snapshots capture times where a wave peak interacts with the accretor (by crossing $y'=0$ at $x'>0$). These frames show a situation of transition between resonances, where the mode pattern is out of phase with the orbit. As a wave peak crosses by the accretor, the ratio of positively to negatively torquing material changes. Comparing the separations of each frame to Figure \ref{fig:tdecay} reveals that the orbital decay first slows (left and center panels) then re-accelerates (right panel) due to gravitational interaction with the passing wave peak.  }
\label{fig:torque}
\end{center}
\end{figure}

However, upon examination of the model snapshots, the explanation for the modulation of the orbital decay rate seen in Figure \ref{fig:tdecay} becomes more clear. Because the binary separation is small, and the mode amplitudes become large, the accretor begins interacting strongly with the perturbed stellar density distribution. Figure \ref{fig:torque} examines the torque density on the orbit from donor-star material in the orbital plane. Green-colored material accelerates the binary's orbital motion (positive torque density), while pink-colored material decelerates the orbit (negative torque density). Imbalance of positively and negatively-torquing material generates a net torque that drives the orbital evolution.  The frames of Figure \ref{fig:torque} are chosen to show a moment when the accretor passes an oscillatory wave peak. While approaching the wave peak the gravitational pull accelerates the orbital motion (left and center panels), and slows the orbital decay. After passing the wave peak, the additional lagging gas density distribution enhances the orbital decay (right panel).

\subsection{Mode Dissipation}

\begin{figure}[tbp]
\begin{center}
\includegraphics[width=0.49\textwidth]{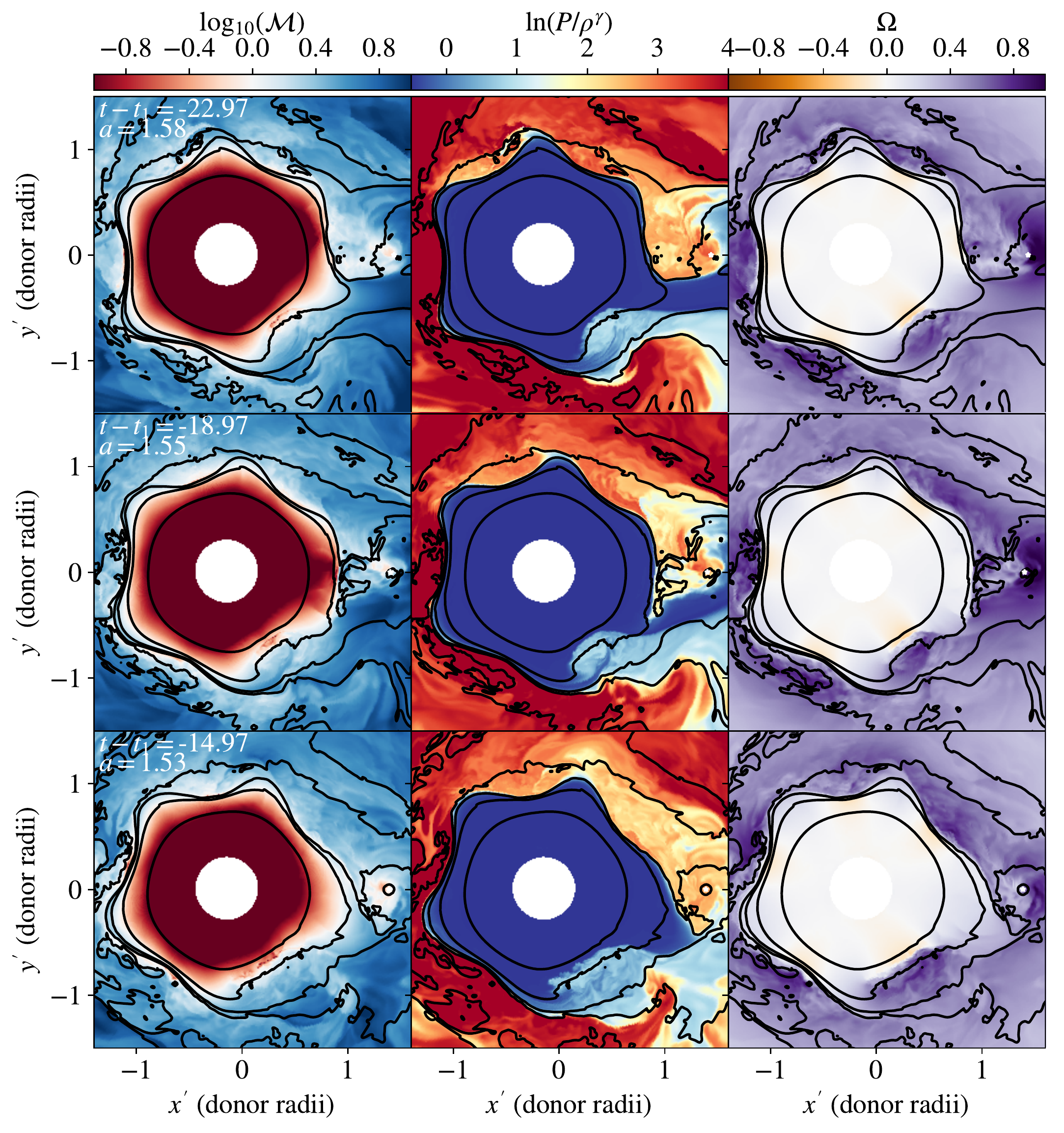}
\caption{ Donor-star envelope Mach number, specific entropy, and angular velocity, with velocities evaluated in the frame of the donor. Rows show the same three times as Figure \ref{fig:torque}, and contours are from $\log_{10}(\rho)=-5$ to $\log_{10}(\rho)=-1$ in one dex intervals.  Supersonic oscillatory velocities are present near wave peaks. When waves crest and break (as in the wave in the lower-right of the donor star in these panels), shocks appear in the gas. This shocking raises the specific entropy of material, as seen in the center panel, creating a surface boundary layer of higher-entropy material on top of the donor's constant-entropy envelope. Wave breaking also redistributes angular momentum, bringing the surface layer toward corotation with the binary orbit ($\Omega_{\rm orb}\approx0.4$).  }
\label{fig:dissipation}
\end{center}
\end{figure}

Oscillatory modes reach large amplitudes in our models, and transition between different modes as the orbit decays. Here we briefly examine dissipation associated with the evolving wave pattern. In our particular scenario, the small mode energies and angular momenta computed in Section \ref{sec:modeorbit} imply that even if the modes were entirely dissipative they would not be expected to significantly disrupt or spin up the donor's envelope. 

Figure \ref{fig:dissipation} examines frames from the time of transition between an $l,m=6$ mode dominating to an $l,m=5$ mode dominating. Each row in Figure \ref{fig:dissipation} shows slices through the orbital plane of one snapshot, with columns showing gas Mach number, specific entropy, and angular velocity (with all velocities measured relative to the donor core). Contours trace 1 dex intervals of density between $\rho = 10^{-5}$ and $\rho=10^{-1}$.

The left panels of Figure \ref{fig:dissipation}, in which the gas Mach number is plotted, show that oscillatory motions are supersonic, ${\cal M} > 1$, near the donor-star limb. Mode pattern velocities are highest here, and the  sound speed is lowest at the stellar limb.  Supersonic motion implies that wave motions will steepen into shocks, particularly where waves crest and break. The influence of these shocks can be traced by examination of the gas specific entropy. The donor's unperturbed envelope structure has constant specific entropy, and shocks raise the entropy of gas in surface layers along the limb of the donor. This effect is most obvious in the dramatically breaking wave seen in the lower right of the donor star in Figure \ref{fig:dissipation}. 

Finally, the shock-heated boundary layer that forms around the donor is pulled from its initial asynchronous state, $\Omega = v_\phi / R = 0$, toward corotation with the orbit, $\Omega_{\rm orb}\approx0.4$.  Because the higher-entropy shock-heated boundary layer material would extend to large scale height in the absence of the accretor's gravity, it dramatically overflows the donor's Roche lobe and is pulled toward the accretor then expelled from the binary system. Therefore, the fact that the system is already Roche lobe overflowing implies that perturbed material is removed from the donor rather than accumulating in a steady-state surface layer. A secondary consequence is that gas that experiences dissipative evolution that pulls it toward corotation is lost from the donor, implying that the donor's average spin rate does not evolve significantly as the binary tightens (as can, for example, be observed in the right-hand panels of Figure \ref{fig:dissipation}).

\subsection{Comments on Potential for Observability}
Although the waves in our model coalescing binary systems have very high amplitudes, our ability to unambiguously detect their presence is not guaranteed. Here we discuss several potentially observable consequences of resonantly-excited modes. 

Photometric variability will be a consequence of the non-radial oscillations excited in our model systems. We might expect this signature to be particularly observable in situations where the binary is not vigorously losing mass, such that the donor-star surface still remains observable. While non-radial stellar oscillations are routinely observed and used for astroseismic analyses, the typical modes are low azimuthal order. Higher order modes suffer increasing geometric cancelation when the distant star is observed as a point source. We consider two possible avenues for photometric detection --  direct detection of higher-azimuthal order oscillations and modification of ellipsoidal variations.

The photometric amplitude at wavelength $\lambda$ of an isolated oscillatory mode can be approximated as
\beq
A_\lambda \approx {\rm abs}\left[ \varepsilon b_l^\lambda (2+l)(1-l) \right],
\eeq
where $\epsilon$ is the wave's radial displacement amplitude $\delta R/R = \varepsilon Y_{lm}(\theta,\phi)$, $b_l^\lambda$ is a stellar disk averaging factor, and $(2+l)(1-l)$ is the geometric influence on the observed flux. This expression is reasonable for $l\gtrsim 6$, where the geometric modulation of stellar flux is dominant over temperature or surface gravity modulations \citep{2002A&A...392..151D}. We also adopt the approximation of a system viewed edge on. For the full expression, see \citet[equation 17]{2002A&A...392..151D}. The disk averaging factor  decreases with increasing $l$, but is approximately compensated by $(2+l)(1-l)$. We arrive, very roughly, at 
$A_\lambda \approx 0.1 \varepsilon$, for even $l$, in the range $l=6$ to $l=12$ (with factor of a few variation depending on $\lambda$). Odd $l$s suffer greater disk averaging and are a factor of nearly 10 further suppressed \citep[see][Fig. 2]{2002A&A...392..151D}. 

Given the relatively large fractional amplitudes, $\varepsilon$, of the tidally-driven waves that arise in our models, photometric detection of similar waves in real stars would be possible, with $A_\lambda \approx 0.01$ for an even-$l$ mode with radial displacement $\epsilon = 0.1$. For a resonant mode, the frequency is set by the resonance condition. The associated photometric variation imprint on the light curve of a binary as a higher frequency modulation with frequency $m \Omega_{\rm orb}$. Space-based facilities like {\it Kepler} and {\it TESS}, can achieve better than milli-magnitude precision for bright stars, implying the ability to measure or constrain an oscillatory modes with amplitude $\varepsilon \gtrsim 0.01$. 

Were high-order oscillations themselves unobservable, it might still be possible to constrain a departure from the expectation of a system exhibiting an equilibrium tide.  In a synchronous system, the projection of the Roche contours of equipotential result in ``ellipsoidal" variations in the light curve due to the projection of the star's $l,m=2$ distortion. Because the bulk of oscillatory power in our models is not in the quadrupole distortion, we expect deviations from the baseline prediction for a synchronously-rotating system exhibiting an equilibrium tidal distortion. At a given orbital period, we expect less amplitude in the quadrupole mode when the binary is asynchronous compared to synchronous.  If the other parameters of the binary system were otherwise constrained (masses, separation, stellar radii -- as is, for example, possible in an eclipsing binary with radial velocity data), it might be plausible to confirm such a deviation.

Other possible avenues for detectability include spectral signatures of atmospheric disturbance by waves, either in the form of line profiles modulated by the surface radial velocity variations \citep[e.g.][]{2012MNRAS.422.1761A}, or emission associated with shock-heated gas. However, atmospheric signatures would only be visible in the scenario of a non-mass-transferring binary. Finally, we note that the non-linear modulation of the orbital decay rate, noted in Section \ref{sec:modeorbit}, might be traceable in binary systems which are seen as eclipsing binaries prior to coalescence. A challenge in the application of this method would be an accurate determination of the binary orbital period over a relatively short duration. Non-linear mode-orbit interactions occur over a timescale less than or similar to an orbital period. This implies that such a measurement would require constraining the orbit of a single cycle, which is generally much less accurate than if a longer time baseline can be accrued and requires high-cadence observations. 

A known binary which may serve as a test case for some of these ideas is V1309 Sco, which underwent an outburst in 2008 \citep{2010A&A...516A.108M} but was serendipitously monitored in the years prior and was shown to be an eclipsing binary with decreasing orbital period \citep{2011A&A...528A.114T}. As such, this system presents the strongest known evidence for a stellar-coalescence driven outburst. Constraints from the eclipsing light curve suggest a relatively asymmetric mass ratio $q\approx 0.1$ \citep{2016RAA....16...68Z}, and it has, therefore, been suggested that the system was driven into merger by asynchronous donor rotation and the Darwin tidal instability \citep{2014ApJ...786...39N}. The pre-outburst light curve, from OGLE, is strongly affected by mass loss from the binary \citep{2014ApJ...788...22P,2017ApJ...850...59P}. However, \citet{2011A&A...528A.114T} state that the typical photometric precision is 0.01~mag, implying that a search for oscillations at integer multiples of the orbital frequency with amplitude $\varepsilon \gtrsim 0.1$  is possible for this source.

\section{Conclusion}\label{sec:conclusion}
We  analyze models of binary coalescence in which the donor-star is asynchronously rotating and show that high-azimuthal order oscillatory modes of the stellar envelope are resonantly excited, making the equatorial structure of the star effectively a polygram. A few key findings of our study are:

\begin{enumerate}

\item As the orbital separation shrinks in the binary's trend toward coalescence, the system sweeps through resonance with different oscillatory modes. At the largest separations, high-order (high $m$) modes are excited, and the resonant-mode order decreases as the orbit tightens (Figures \ref{fig:resonancecrossing} and \ref{fig:slice}).

\item In a rapidly-decaying binary system, the peak-excitation of a mode lags the crossing of the maximal resonance separation (Figure \ref{fig:powersep}), but the oscillatory amplitude predicted by linear theory is otherwise accurate to within a factor of two even as the modes attain quite large amplitudes (Figure \ref{fig:vrms}). 

\item Stellar rotation affects mode excitation,  as explored in Section \ref{sec:rotation}. At a given binary separation higher order modes are excited as the stellar rotation rate increases (see Figure \ref{fig:rot}).

\item Directly observing the photometric imprint of high azimuthal order modes is possible but challenging. We estimate that the photometric amplitude of even-$l$ modes is $A_\lambda \approx 0.1 \varepsilon$ for high-order modes. Therefore, milli-magnitude level photometry could constrain the presence of any modes with fractional amplitude greater than 1\%, $\varepsilon\gtrsim 0.01$. Modes in resonance would exhibit a frequency that is an integer multiple of the orbital frequency. 

\end{enumerate}

We have considered a restricted scenario of a donor star with an isentropic envelope transferring mass and coalescing with a lower-mass companion. However, the qualitative features of our results should be applicable to a wide range of astrophysical systems. Non-isentropic structures, such as those of stars with radiative envelopes, also support $g$-modes, which may be resonantly excited \citep[e.g.][]{1994MNRAS.270..611L,2011MNRAS.412.1331F,2012MNRAS.420.3126F,2018MNRAS.476..482V}. Our initial testing via simulation models indicates that high azimuthal order modes may be selectively excited in these cases as well, and the linear model described in section \ref{sec:linear} provides a framework for exploring this more fully.   Other astrophysical scenarios of possible interest and applicability for our findings include close-in planetary perturbers such as Hot Jupiters and the plunge toward merger in double neutron star systems (For example, see Fig. 24 of \citet{2017RPPh...80i6901B} 
and \citet{2004MNRAS.352.1089S}).

\acknowledgments
We gratefully acknowledge helpful discussions with A. Dupree, J. Fuller, T. Kaminski, J. Ostriker, S. Schr\o der, and S. Tremaine. We thank R. Townsend for advice on modeling our stellar structure in GYRE. Finally, we thank E.C. Ostriker for invaluable assistance in the early stages of this project. 

M.M. is grateful for support for this work provided by NASA through Einstein Postdoctoral Fellowship grant number PF6-170169 awarded by the Chandra X-ray Center, which is operated by the Smithsonian Astrophysical Observatory for NASA under contract NAS8-03060. 
Support for program \#14574 was provided by NASA through a grant from the Space Telescope Science Institute, which is operated by the Association of Universities for Research in Astronomy, Inc., under NASA contract NAS 5-26555.
M.V. acknowledges support from the NASA Earth and Space Sciences Fellowship in Astrophysics.
DL acknowledges support from 
the NSF
grant AST1715246 and NASA grant NNX14AP31G.
Resources supporting this work were provided by the NASA High-End Computing (HEC) Program through the NASA Advanced Supercomputing (NAS) Division at Ames Research Center.

\software{  Athena++, Stone et al. (in preparation) \url{http://princetonuniversity.github.io/athena}, Astropy \citep{2013A&A...558A..33A}, GYRE \citep{2013MNRAS.435.3406T,2018MNRAS.475..879T}, shtools \citep{shtools} }

\bibliographystyle{aasjournal}

\end{document}